\documentclass[fleqn,usenatbib]{mnras}

\usepackage{newtxtext,newtxmath}

\usepackage{sidecap}

\usepackage[T1]{fontenc}

\DeclareRobustCommand{\VAN}[3]{#2}
\let\VANthebibliography\thebibliography
\def\thebibliography{\DeclareRobustCommand{\VAN}[3]{##3}\VANthebibliography}

\usepackage[normalem]{ulem}

\usepackage{graphicx, caption, subcaption}	
\usepackage{amsmath}	

\usepackage[dvipsnames]{xcolor}

\title[Environmental Dependence of the MZR]{The Environmental Dependence of the Stellar Mass - Gas Metallicity Relation in Horizon Run 5}

\author[A.R. Rowntree et al.]{Aaron R. Rowntree$^{1}$ \thanks{E-mail: a.rowntree-2018@hull.ac.uk (ARR)}, Ankit Singh$^{2}$, Fiorenzo Vincenzo$^{1}$, Brad K. Gibson$^{3}$, C\'{e}line Gouin$^{4}$, \newauthor Daniela Gal\'{a}rraga-Espinosa$^{5}$, Jaehyun Lee$^{7}$, Juhan Kim$^{2}$, Clotilde Laigle$^{6}$, Changbom Park$^{2}$, \newauthor Christophe Pichon$^{2,6,10}$, Gareth Few$^{1}$, Sungwook E. Hong$^{7,8}$, Yonghwi Kim$^{9}$ 
\\ ~ \\
$^{1}$E.~A. Milne Centre for Astrophysics, University of Hull, Hull, HU6 7RX, UK\\
$^{2}$Korea Institute for Advanced Study (KIAS), 85 Hoegiro, Dongdaemun-gu, Seoul 02455, Republic of Korea\\
$^{3}$Woodmansey Primary School, Hull Road, Woodmansey, HU17 0TH, UK\\
$^{4}$Universit\'{e} Paris-Saclay, CNRS, Institut d’Astrophysique Spatiale, 91405 Orsay, France\\
$^{5}$Max-Planck-Institut f\"{u}r Astrophysik, Karl-Schwarzschild-Str. 1, D-85748 Garching, Germany\\
$^{6}$Institut d’Astrophysique de Paris, UMR 7095, CNRS, and Sorbonne Universit\'{e}, 98 bis Boulevard Arago, F-75014 Paris, France \\
$^{7}$ Korea Astronomy and Space Science Institute, 776 Daedeok-daero, Yuseong-gu, Daejeon 34055, Republic of Korea \\
$^{8}$ Astronomy Campus, University of Science and Technology, 776 Daedeok-daero, Yuseong-gu, Daejeon 34055, Republic of Korea \\
$^{9}$ Korea Institute of Science and Technology Information, 245 Daehak-ro, Yuseong-gu, Daejeon, 34141, Korea \\
$^{10}$ IPhT, DRF-INP, UMR 3680, CEA, L'Orme des Merisiers, B\^at 774, 91191 Gif-sur-Yvette, France\\
}

\date{Accepted 2024 May 31; Received 2024 May 16; in original form 2024 April 10}
\pubyear{2024}

\begin{document}

\maketitle

\begin{abstract}
    Metallicity offers a unique window into the baryonic history of the cosmos, being instrumental in probing evolutionary processes in galaxies between different cosmic environments. We aim to quantify the contribution of these environments to the scatter in the mass-metallicity relation (MZR) of galaxies. By analysing the galaxy distribution within the cosmic skeleton of the Horizon Run 5 cosmological hydrodynamical simulation at redshift $z = 0.625$, computed using a careful calibration of the {\tt T-ReX} filament finder, we identify galaxies within three main environments: nodes, filaments and voids. We also classify galaxies based on the dynamical state of the clusters and the length of the filaments in which they reside. We find that the cosmic environment significantly contributes to the scatter in the MZR; in particular, both the gas metallicity and its average relative standard deviation increase when considering denser large-scale environments. 
    The difference in the average metallicity between galaxies within relaxed and unrelaxed clusters is $\approx 0.1 \text{ dex}$, with both populations displaying positive residuals, $\delta Z_{g}$, from the averaged MZR. Moreover, the difference in metallicity between node and void galaxies accounts for $\approx 0.14 \, \text{dex}$ in the scatter of the MZR at stellar mass $M_{\star} \approx 10^{9.35}\,\text{M}_{\sun}$. Finally, both the average [O/Fe] in the gas and the galaxy gas fraction decrease when moving to higher large-scale densities in the simulation, suggesting that the cores of cosmic environments host -- on average -- older and more massive galaxies, whose enrichment is affected by a larger number of Type Ia Supernova events.
\end{abstract}


\section{Introduction}
\label{sec:intro}
The distribution of matter on the largest scales of the Universe is a complex entanglement of dark matter (DM) and baryonic matter in the form of gas, dust and stars. 
By using galaxies as tracers of the matter distribution in the cosmos, large-scale structures emerge, forming complex patterns that determine the so-called `cosmic web' (e.g., \citealt{Peebles1980, deLapperant1986, Shandarin1989, Geller1989, Bond1996}).
Gravity not only governs the formation of the overall cosmic web but also dictates how matter moves and amalgamates to form the individual structures themselves. Over time, substructures in low-density regions are attracted to high-density regions, slowly increasing the concentration of mass within them \citep{Zeldovich1970, Bardeen1986}. This amplifies the density contrast between areas in the cosmos and the more matter that accumulates, the clearer these large build-ups are and the deeper their gravitational potential wells become, giving rise to unique cosmic environments. \textit{(i)} Nodes house up to thousands of galaxies, giving rise to the highest density regions hosting large galaxy clusters \citep{Gregory1978}. \textit{(ii)} Filaments act as the rivers connecting the nodes to one another, facilitating the flow of matter throughout the cosmos \citep{Bond1996}, and \textit{(iii)} voids are the almost empty spaces outside of the nodes and filaments that are left behind (e.g., \citealt{Kirshner1987}).

%Environment and galaxy properties 

A number of studies showed that galaxy properties such as star formation rate (SFR) \citep{Gavazzi2002, Porter2008, Peng2010, Haines2011, 2016Alpaslan, 2016Mart, Mahajan2018, Gallazzi2021}, gas-fraction \citep{Hasan2023}, morphology \citep{2017Kuutma} and metallicity \citep{Shields1991, Henry1992, Donnan2022}, are affected by the cosmic environment that they exist within. Further studies into these environmental dependencies explained what is causing these relationships. For example, high-density regions have a higher frequency of galaxy mergers \citep{LHullier2012} and also possess the optimal conditions to cause tidal stripping \citep{Jhee2022}, both leading to dramatic changes in the properties of the galaxies involved. Using the cosmological-hydrodynamical simulation IllustrisTNG-100 \citep{Nelson2019}, \citet{Gupta2018} reported a $\sim 0.05$ dex enhancement in metallicity at $z \leq 1.0$ for galaxies that have fallen into clusters.

% MZR

Metallicity is of key importance when studying galaxy formation. Both stellar and gas metallicities have been studied in excess throughout the history of astronomy, and therefore it is important to distinguish between them. Stellar metallicity is a parameter that shows the integrated metallicity evolution up until the time at which a star forms, snapshotting the metallicity at that instant. Gas metallicity on the other hand instead, is a snapshot of the metallicity at the present moment, and this is the property that we choose to study in this work. Several properties of a galaxy are either directly linked or show correlation with metallicity: inflows, feedback processes such as outflows in the form of stellar winds, active galactic nuclei (AGNs) and supernova (SN) explosions, stellar mass, SFR, gas-fraction and environment. Metallicity, therefore, provides a clear window into baryonic evolution in galaxies. 
When discussing metallicity, one must also consider the mass-metallicity relation (hereafter MZR) and what governs the variation between stellar mass and metallicity in a galaxy \citep{Tremonti2004}. The MZR has been extensively studied in the local Universe \citep{1979Lequeux, Tremonti2004} and at higher redshifts \citep{2005Savaglio, 2006Maier, 2008Maiolino, 2012Foster, 2013Zahid, 2013Moller}. The dependence of the MZR on the environment has also been investigated by several studies in the literature \citep{Thomas2005, 2012Yates, 2013Pilyugin, 2013Sanchez,2014Peng}. For example, early studies into metallicity and environment reported higher chemical enrichment for Virgo group galaxies compared to the field (e.g., see \citealt{Shields1991, Henry1992, Skillman1996}). At redshift, $z \approx 0.53$, in the COSMOS field, \citet{Darvish2015} spectroscopically identified a large filament, reporting a $\sim0.10$-$0.15$ dex metal enrichment for galaxies associated with the filament. By analysing Sloan Digital Sky Survey (SDSS) data to explore the impact of local density and cluster membership on the MZR, \citet{2009Ellison} reported higher metallicities for cluster galaxies independent of galaxy size and cluster properties. \citet{2017Wu} also studied the MZR relation as a function of local density in SDSS, reporting a slight dependence of MZR relation on the environment.

Over the last two decades, many efforts have been made to explain the vertical scatter in the MZR, which is consistently larger than measurement uncertainty. Observationally,  \citet{Choi2014} measured the stellar MZR at $z = 0.7$, finding different metallicities for individual galaxies when compared to the total measured MZR at the same redshift, further indicating a large scatter within the population. 
\citet{2015Tran} conducted a spectroscopic survey of the cluster XMM-LSS J02182-05102 at $z\sim1.6$ in optical and infrared wavelengths. They reported cluster galaxies at $z\sim1.6$ to lie below the MZR relation of local galaxies, finding no environmental dependence of how the gas metallicity changes with stellar mass. Several studies have used cosmological-hydrodynamical simulations to explore the physical dependencies of the scattering in the MZR. For example, \citet{Torrey2019} found the scatter in the MZR to strongly correlate with the gas mass and the SFR of the system. Using a semi-analytical approach \citet{Delucia2020} demonstrated metallicity dependence on the galaxy's gas-accretion history and using the {\tt{EAGLE}} simulations, \citet{vanLoon2021} further specified clear dependencies between the scatter and gas-fraction, inflow rate and outflow rate. They report that although these variables are closely coupled, each does have an independent contribution with gas-fraction itself, completely determining the relation between residual metallicity and residual specific star formation rate (sSFR), also see \citealt{Chen2022} for an observational perspective. More recently \citet{Wang2023} analysed the gas-MZR over a redshift range of $0-2$, finding that, at high redshift, the accretion of low-metallicity gas was responsible for any environmental dependencies, whereas at low redshift, AGN feedback played a more crucial role. AGN feedback continues to show a large importance in \citet{Yang2024} for high stellar mass galaxies, at which gas accretion is shown to no longer be the largest cause of scatter in the MZR with scatter from AGN feedback being more prominent. \citet{Donnan2022} links the MZR back into the field of large-scale structure, clearly connecting the MZR and cosmic environments. They reported higher levels of chemical enrichment for galaxies closer to the spine of filaments and lower enrichment for field galaxies, demonstrating a significant variation between the two populations in the context of the MZR.

It is difficult to acquire accurate spectroscopy to measure metallicity alongside data probing large-scale structures however, with the advent of Dark Energy Survey Instrument \citep[DESI; ][]{DESI2023}, we can expect to be able to access good quality spectroscopic data in the coming years. It is, therefore, crucial to unravel the connections between metallicity and the cosmic web further using the currently available tools to help provide insight for future studies and build upon the work that has already been established.

This work presents the study of the scatter in the gas MZR in different environments, using the cosmological-hydrodynamical simulation Horizon Run 5 (HR5) \citep{Lee2020}. With a large box size of 1 $\text{cGpc}^{3}$ and 1 kpc resolution, HR5 can capture large modes of structure formation whilst providing accurate galaxy properties and is ideal for the present study. Through this analysis, we will build on our understanding of the cosmic hierarchy, how galaxies evolve, and how much the global environment affects galaxy metallicity and the active processes that influence it. Our results hope to clearly define the portion of the scatter in the MZR that the global environment can account for and to provide another perspective on existing relationships to help underpin future work using upcoming data releases.

Our work is structured as follows. Section \ref{sec:Method} presents the HR5 simulation, our methodology to calculate the skeleton, and the galaxy catalogues used for our analysis, along with our environmental classification. Our results are presented in Section \ref{sec:results} and discussed in Section \ref{sec:disc}. Finally, we conclude with our major findings in Section \ref{sec:Concl}.

\section{Method}
\label{sec:Method}

\subsection{Simulation Data}
\label{sec:sim_data}

\subsubsection{Horizon Run 5}
\label{sec:hr5}

Horizon Run 5 (HR5) is one of the latest cosmological hydrodynamical simulations available \citep{Lee2020}. The simulation was run using a hybrid {\tt MPI-OpenMP} version of the {\tt RAMSES} adaptive mesh refinement code \citep{Teyssier2002} to maximize computing efficiency using modern architectures. It follows the formation and evolution of galaxies and cosmic structures up to redshift $z=0.625$ within a box that covers a volume of $1.049 \times 1.049 \times 1.049$ $\text{cGpc}^{3}$ in co-moving units, whilst achieving a resolution of 1 kpc within the higher-resolution, zoom-in region with a cuboidal volume of $1049 \times 119 \times 127$ $\text{cMpc}^{3}$. HR5 employs a large range of sub-grid physics, including metallicity-dependant radiative cooling \citep{Dalgarno1972, Sutherland1993}, UV background heating \citep{Haardt1996}, star formation activity \citep{Rasera2006}, SN \citep{Dubois2008} and AGN feedback \citep{Dubois2012}. The simulation also bases its overall chemical enrichment model on \citet{Few2012} that includes massive stars dying as core-collapse SN and exploding white dwarfs giving rise to Type Ia SNe, allowing us to track the evolution of the oxygen and iron abundances in the interstellar medium (ISM) and stellar populations within galaxies. HR5 uses the following cosmological parameters for a standard $\Lambda$ cold dark-matter ($\Lambda$CDM) Universe: $\Omega_{0}=0.3$, $\Omega_{\Lambda}=0.7$, $\Omega_{b} =0.047$, $\sigma_{8} = 0.816$, and $H_{0}= 100\times h_{0}=68.4\,\text{km}\,\text{s}^{-1}\,\text{Mpc}^{-1}$, that are compatible with the Planck data \citep{Plank2016}.

\subsubsection{Galaxy Catalogue and Selection Criteria}
\label{sec:galcat}

To populate its galaxy and halo catalogues, HR5 makes use of an extended friends-of-friends (FoF) algorithm that identifies virialized halos by creating a chain of linkages between DM, stellar and black hole (BH) particles. These galaxy groups are then passed to the PSB-Based Galaxy Finder, {\tt PGalF}, which is based on the original Physically Self-Bound (PSB) algorithm \citep{Kim2006}. {\tt PGalF} identifies a subhalo by searching for peaks at each particle position in the stellar mass density field. Each peak's core and possible non-core particles are extracted, creating a galaxy candidate. A membership decision for each particle is then carried out, checking whether it lies within the tidal boundary of the galaxy. The total energy of the particles within the boundary is checked to ensure a valid result, and a successful candidate is added to the catalogue (see \citealt{Lee2020, Kim2023} for more details).

Our analysis is based on the galaxy catalogue for the last snapshot at redshift $z=0.625$. For this study, we are specifically interested in total stellar mass, total gas mass, average gas-phase metallicity, SFR, and positions in cartesian space. Note that all galaxy gas properties are computed within a radius of $30$ kpc from the identified sub-halo center to extract only the information associated with the interstellar medium.

The raw catalogue requires certain quality cuts before we can begin our analysis. We begin by removing all galaxies flagged as `impure' in the catalogue; these galaxies lie close to large, low-resolution regions, causing their properties to be unreliable for our analysis. Within this first cut, we only consider galaxies with stellar mass $M_{\star}> 2 \times 10^{9}\,\text{M}_{\sun}$. In HR5, stellar particles have a minimum mass of $2 \times 10^{6}M_{\sun}$ meaning galaxies with $M_{\star}> 2 \times 10^{9}\,\text{M}_{\sun}$ have at least 1000 stellar particles, making them high enough resolution for this study. These two cuts give us a total of $158,094$ galaxies in the final snapshot, irrespective of their position in the large-scale structure in HR5, constituting what will be referred to as $S_{\rm all}$. One final cut must be made to focus on the filamentary structure in HR5, removing the galaxies associated with large clusters in the simulation \citep{Chen2017}. We do this by identifying all clusters with total mass $M_{\text{cluster}} > 10^{13}\,\text{M}_{\sun}$ and then removing all galaxies that lie at distances within $2\times R_{200}$ from the cluster centre, where $R_{200}$ corresponds to the radius at which the mean matter density is $200$ times the critical density of the Universe, similar to the cluster cut seen in \citet{Galarraga2023}. The cut-off mass $M_{\text{cluster,cut}} = 10^{13}\,\text{M}_{\sun}$ effectively targets large clusters at $z=0.625$, which lie at the nodes of the large-scale structure in HR5. After this final removal, we are left with a total of $69,214$ galaxies referred to as $S_{\rm other}$.

\subsection{Cosmic Skeleton Computation}
\label{sec:skel}

To establish any connection between metallicity and a galaxy's global environment, one must be able to define the global environment of a galaxy. In recent years, this has been achieved through the use of filament finding algorithms such as {\tt DisPerSE} \citep{Sousbie2011}, {\tt Nexus} \citep{Cautun2013} and {\tt T-ReX} \citep{Bonnaire2020, Bonnaire2022}. Each algorithm builds the estimated structures through a different methodology, e.g. using the topology of the matter density field \citep{Sousbie2011}, Hessian matrices \citep{Cautun2013}, regularised minimum spanning trees \citep{Bonnaire2020}, and recently machine learning methods \citep{Inoue2022, Awad2023}. For a more detailed comparison between different methodologies of filament classification, we refer the readers to \citet{Libeskind2018}. 

To estimate the large-scale structure in a dataset, the chosen filament finding algorithm must be provided with a matter distribution, be that galaxies, DM, or both. The DM distribution provides the most `true-to-life' tracing of the large-scale structure at a given epoch of the Universe or simulation \citep{Song2021}. Since observationally, it is difficult to retrieve the DM distribution; the galaxies are commonly used as a tracer. Whilst biased and considered less accurate than DM \citep{Laigle2018}, using galaxies to trace the web has been shown to correctly trace the underlying properties of the density field and its geometry \citep{Kraljic2018}. \citet{Zakharova2023} also explicitly showed that filamentary structures identified in the same dataset using different tracers reasonably agree. This provides confidence in using galaxies to trace the cosmic web (e.g. \citealt{Galarraga2020, Gouin2021, Donnan2022, Bulichi2023, Galarraga2023_2}). 

Depending on the proximity to an identified structure, we can explicitly define the cosmic environment associated with a galaxy by using these filament-finding algorithms. Recently \citet{Bonnaire2020} introduced a new filament finder, {\tt T-ReX}. {\tt T-ReX} is designed to detect filamentary structure in observations based on a point distribution. It uses regularized minimum spanning trees and graph theory to estimate the structure. {\tt T-ReX} uses a combination of Gaussian Mixture Models (GMM) to describe the distribution of the tracer particles while graph theory is used to extract a smooth, minimum spanning tree that connects the tracers \citep{Gouin2021}. This methodology succeeds in this context as the structure is a continuous feature, whilst the tracers are an infrequent, noisy sampling of the feature we wish to recover \citep{Bonnaire2020}. In summary, {\tt T-ReX} statistically samples the cosmic structure present in the distribution of tracer particles, outputting a list of edges in space that make up a skeleton-like structure. From here onwards, this list of edges referred to as the `skeleton' can then be used to define the cosmic environments numerically.

To begin our analysis using these catalogues in the context of the large-scale structure in HR5, we must quantify the structure itself. Similar to previous works (e.g., see \citealt{Sarron2019, Galarraga2020, Gouin2021, Donnan2022, Bulichi2023}), we choose to use the galaxies tracer of the structure. Combining this with the {\tt T-ReX} Filament Finder \citep{Bonnaire2020} we can extract an accurate, numerical estimate of the positions of the different environments. For the specific python implementation we utilize, see \citealt{Bonnaire2021}.

One can use multiple methods and cuts to decide on a galaxy selection for skeleton computation. The most common approach is to apply a mass cut in the galaxy catalogue. For example, \citet{Sarron2019, Galarraga2020} restricted their input data to galaxies with total stellar mass $M_{\star} > 10^{9-9.5}\,\text{M}_{\sun}$ to construct their skeleton. Generally, in the literature, this methodology is physically justified because higher stellar mass galaxies tend to be found more frequently within filaments and nodes, so using these as the tracer accurately represents the underlying large-scale structure. This method is also coherent with galaxy selections that are utilised in observational work. However, this method's main limitation is that it only samples part of the stellar mass distribution. Irrelevant of stellar mass, all galaxies are gravitational bodies contributing to the structure. Moreover, galaxies above the mass cut used in the skeleton computation have a different distribution in their `distances to the filament spine' to those below the mass cut. As such, if the mass cut used in the galaxy property analysis is above that employed for the skeleton computation, then the calculated values of this distance value are inconsistent (see Appendix A). This leads to our second approach: randomly sample the $20\%$ of the galaxies across the full stellar mass distribution. The reason we choose to use a $20\%$ sampling rather than the full $100\%$ is outlined in Appendix A; however, in short; using the full distribution leads to a significant overpopulation of short filaments, leading to an incorrect, bimodal PDF in filament length when compared to existing literature, \citep{Galarraga2023_2}. Using a $20\%$ sample also leads to a galaxy number density comparable to that seen after taking a mass cut of $M_{\star} \geq 10^{9}$, yet it maintains the full stellar mass distribution. This leads to a much more comparable, unimodal PDF in filament length, over the correct range. 

This study uses the second methodology outlined above, tuning {\tt T-ReX} to retrieve a viable representation of the cosmic web in HR5. We begin with the original catalogue of $901,985$ galaxies to compute our skeleton. The only galaxy removal implemented before skeleton computation is the removal of the `impure' galaxies found near low-resolution regions, opting not to utilize the common $M_{\star}$ cut as seen in aforementioned studies. This leaves $645,970$ galaxies, and after taking $20\%$ of these, we have $129,194$ galaxies to be utilize in the creation of the skeleton. The main parameters within {\tt T-ReX} are the following \citep{Bonnaire2020}: \textit{(i)} $\Lambda$, that governs the smoothness of the skeleton; \textit{(ii)} $l$, that governs the level of de-noising that occurs in the calculation, and \textit{(iii)} $\sigma$, which alters the variance of the GMMs that are in use. For our catalogue of galaxies at redshift $z=0.625$ in HR5, we use $\Lambda = 5$, $l = 2$ and $\sigma = 2$. 

\begin{figure}
    \centering
    \includegraphics[scale=0.65]{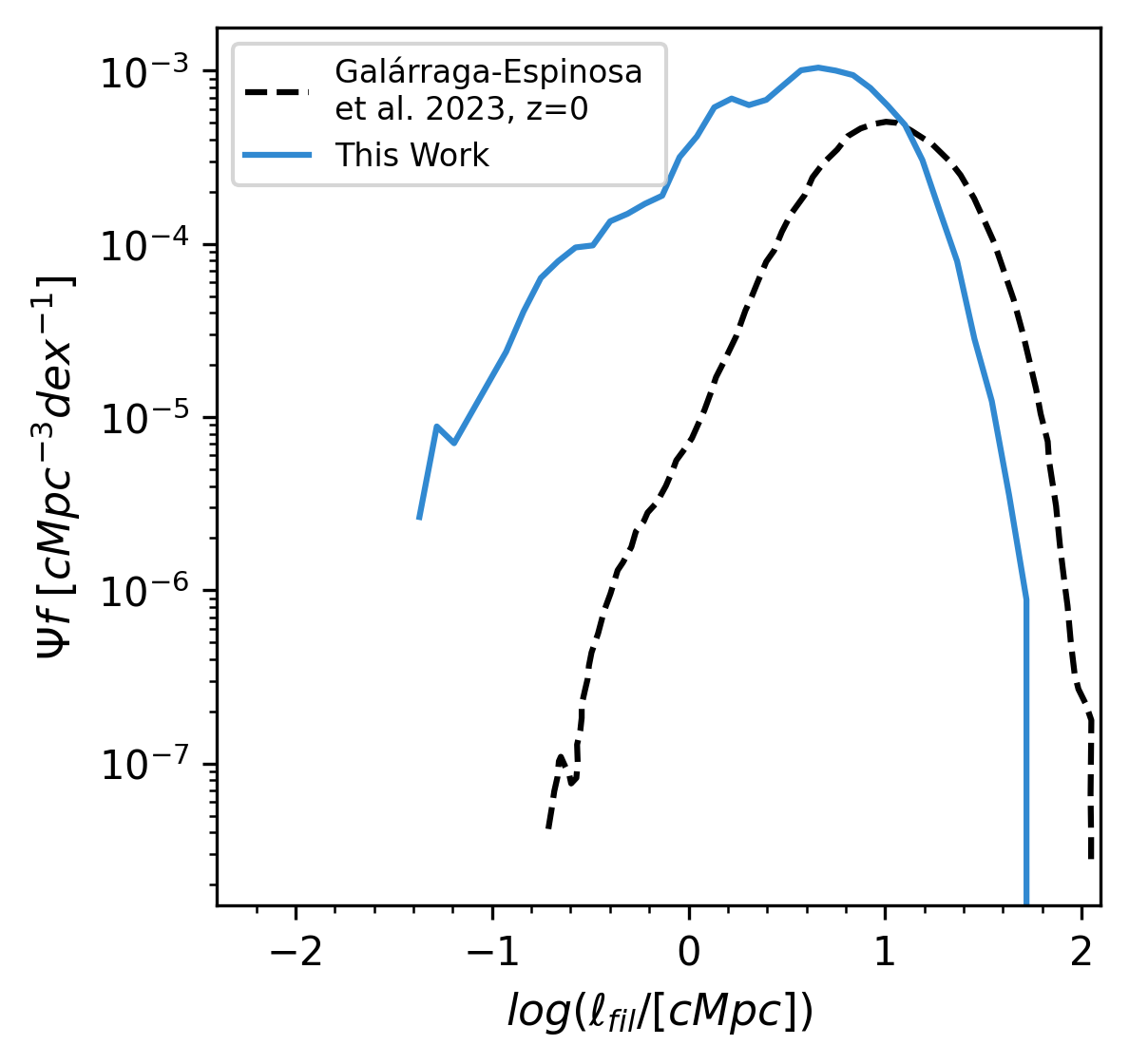}
    \caption{The filament length function, $\Psi_{f}$, for this work, compared to that seen in \citet{Galarraga2023_2}. The solid blue line represents the function associated with this work, whilst the dashed black line presents the function associated with the study we are comparing to.}
    \label{fig:FilLenFunctions}
\end{figure}

To finally assess the consistency of our skeleton with literature, we calculate the filament length function as seen in \citet{Galarraga2023_2}. The filament length function, $\Psi_{f}$, is defined as the number density of filaments in a bin of length re-scaled by the bin width. Fig. \ref{fig:FilLenFunctions} compares our work and the filament length function associated with \citet{Galarraga2023_2}. The similarity in the functions, considering the redshift difference, shows that even with different simulations, filament finders and number densities, we are working in similar spatial scales and as such our skeleton is achieving our goals.
Further explanation of the decision to use the second approach and specific reasoning for parameter choices can be found in Appendix A.

\begin{figure}
    \centering
    \includegraphics[scale=0.65]{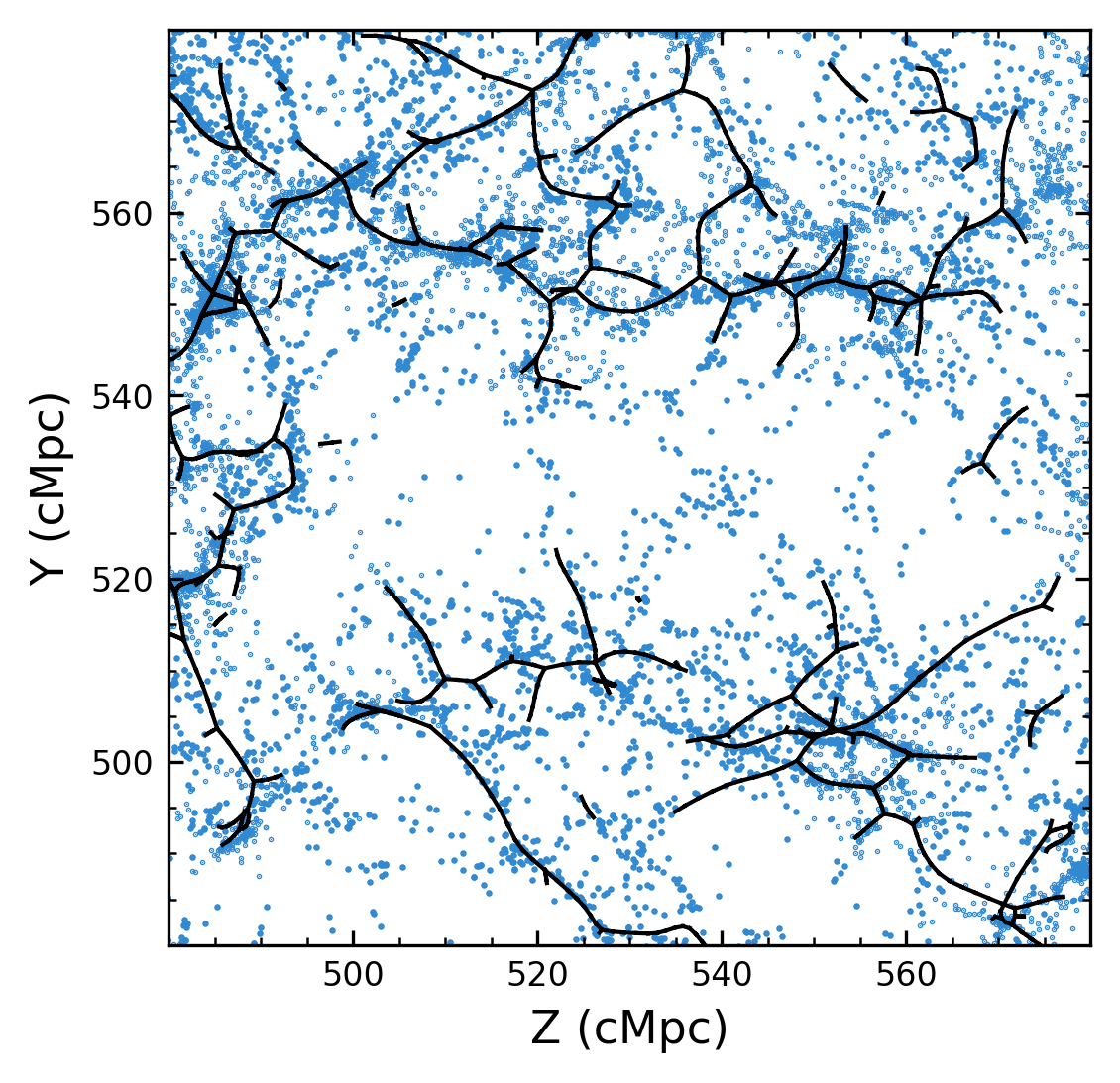}
    \caption{The computed skeleton with {\tt T-ReX} (black solid lines) overlayed onto the galaxy distribution (blue points) in a projected $20\,\text{cMpc}$ slice of HR5. Visually, this demonstrates that the filament finder is able to reasonably identify the filaments in the dataset.}
    \label{fig:2dSliceSkeleton}
\end{figure}

Fig. \ref{fig:2dSliceSkeleton} shows a galaxy distribution within a 2D slice of $20$ cMpc thickness from the zoom-in region of HR5, overlayed with the skeleton identified by {\tt T-ReX}. The skeleton allows us to calculate the perpendicular distance of each galaxy to the nearest filament ($d_{\rm skel}$). To compute $d_{\rm skel}$, we utilize the {\tt radial\_distance\_skeleton} function provided in the {\tt T-ReX} library. Executing this between each galaxy in $S_{\rm other}$ and the skeleton itself gives us each galaxy's perpendicular distance to the nearest edge on the skeleton, $d_{\rm skel}$. In this work we define cluster galaxies or, equivalently, node galaxies, as galaxies that are within $2 \times R_{200}$ of FoF halos with $M_{\rm tot}$ $\geq 10^{13}\,\text{M}_{\sun}$. For this population, we can define $d_{\rm cluster}$ as the distance of an individual galaxy to the center of mass of the FoF halo that it belongs to, where the total mass of the halo is $M_{\rm tot}\geq 10^{13}\,\text{M}_{\sun}$. We note that $d_{\rm cluster}$ is computed by using the nearest neighbour algorithm carried out between $S_{\rm all}$ and the high-mass galaxy cluster dataset. 

\section{Results}
\label{sec:results}
\subsection{Spatial Distribution of Galaxies in the Skeleton}
\label{sec:res_spad}

\begin{figure}
    \centering
    \includegraphics[scale=0.655]{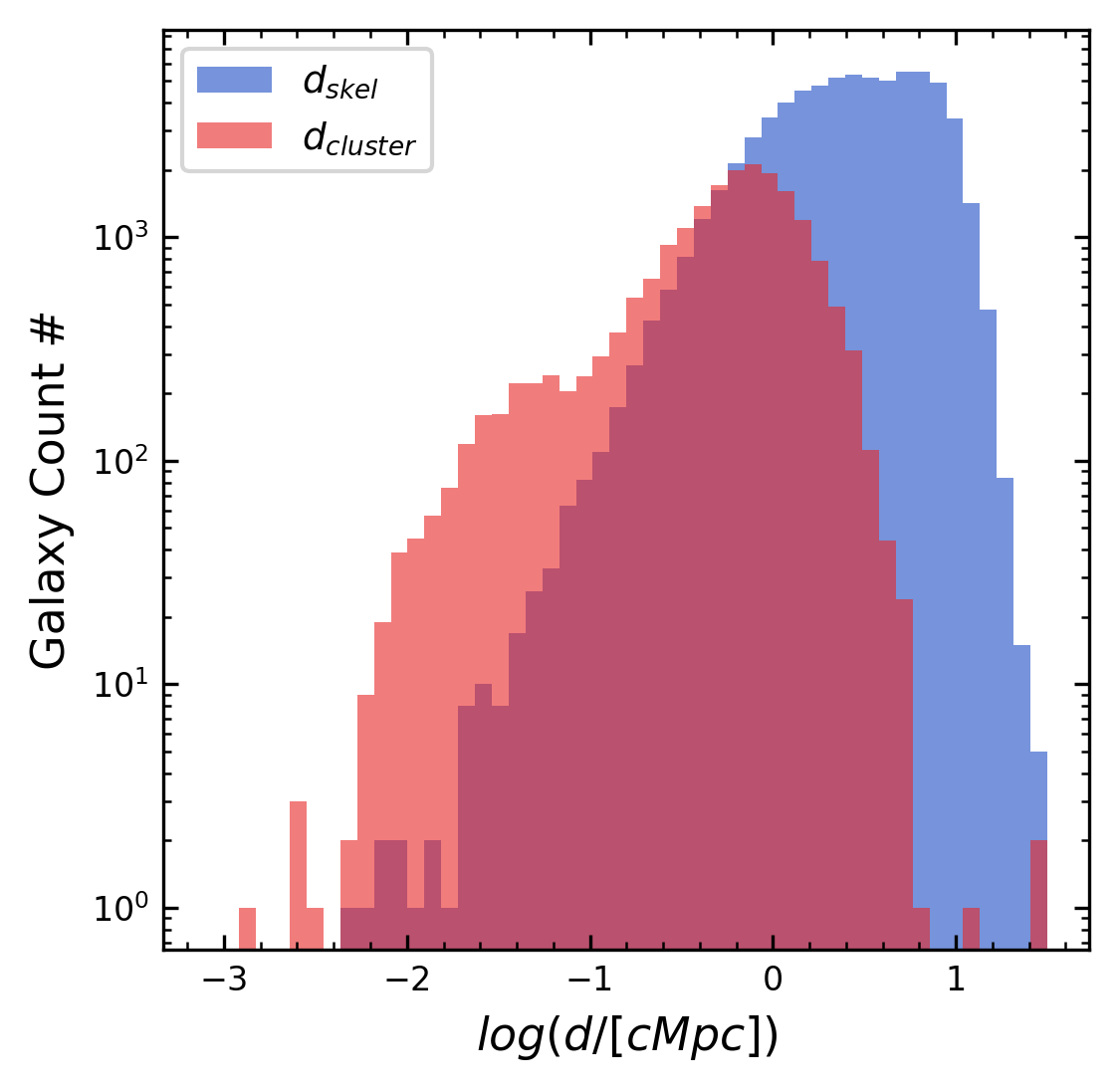}
    \caption{Overlayed 1-D histograms for the distance to the nearest filament, $d_{\rm skel}$, in blue, and for the distance to the nearest galaxy cluster, $d_{\rm cluster}$, in orange. Note that $d_{\rm cluster}$ includes galaxies within $2\times R_{200}$ of each galaxy cluster with $M_{\rm tot} \geq 10^{13}\,\text{M}_{\sun}$, whilst $d_{\rm skel}$ includes the galaxies outside of this cut. Each distance measure is split into 50 equally sized logarithmic bins in the range $-3.1 \leq \log(d/[\text{cMpc}]) \leq 1.1$.}
    \label{fig:FilClustDistributions}
\end{figure}

\begin{figure}
    \centering
    \includegraphics[scale=0.65]{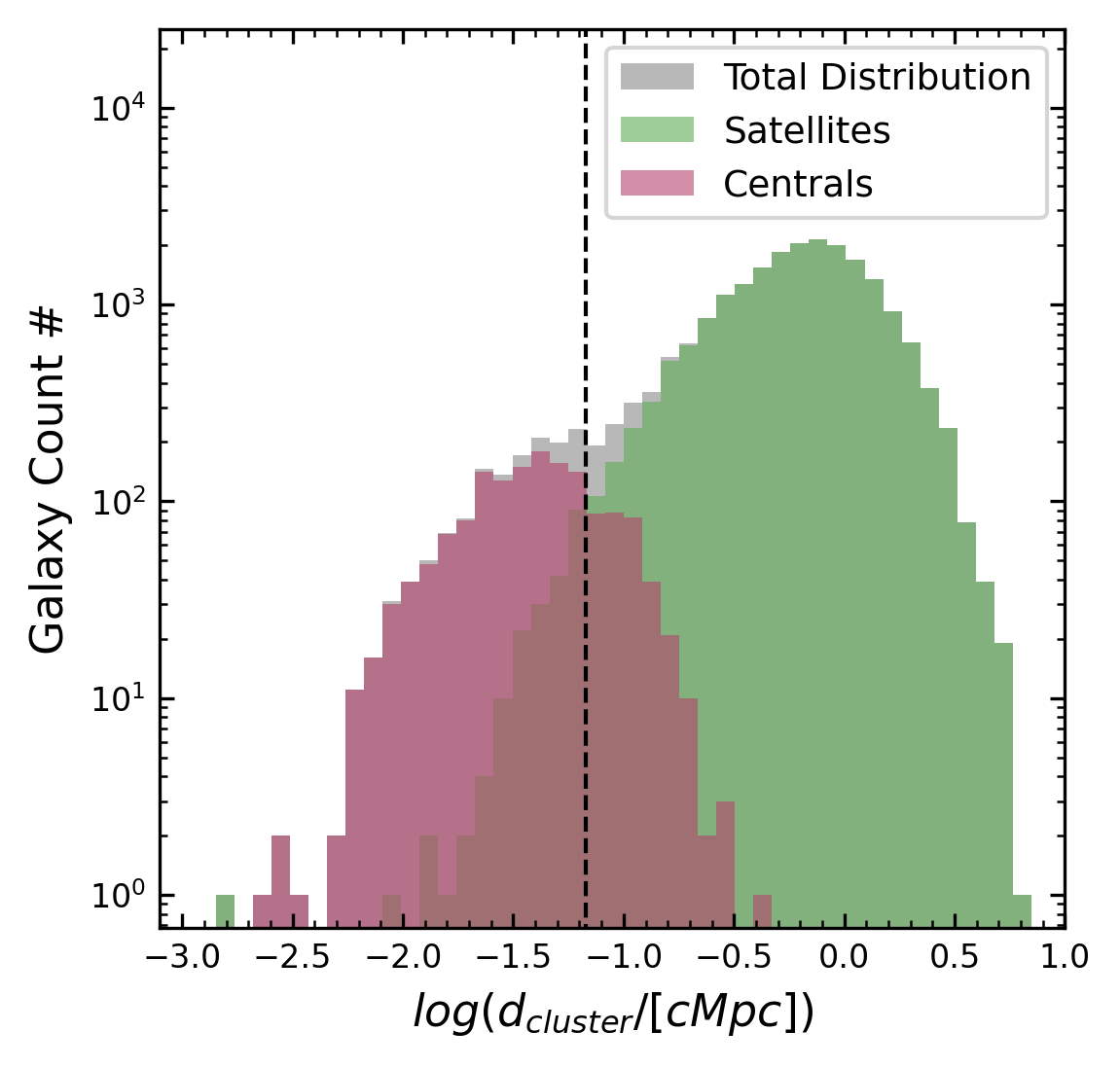}
    \caption{Overlayed 1-D histograms of the distance to the nearest galaxy cluster, $d_{\rm cluster}$, considering only galaxies lying within $2\times R_{200}$ of each identified galaxy cluster. The total distribution of $d_{\rm cluster}$ is shown in grey, the satellite galaxy distribution in blue, and the central galaxy distribution in orange. The vertical black dashed line denotes the distance at which central galaxies become the more prominent population, $d_{\rm cluster}\space = -1.17$ log(cMpc).}
    \label{fig:SatCentDists}
\end{figure}

In Fig. \ref{fig:FilClustDistributions}, the distribution of $d_{\rm skel}$ is compared to the distribution of $d_{\rm cluster}$ for each galaxy. The distribution of $d_{\rm cluster}$ is vaguely bimodal, showing a first 'bump' at $\log(d_{\rm cluster}/\text{cMpc})=-1.4$ and a second peak at $\log(d_{\rm cluster}/\text{cMpc})=-0.2$. The presence of the first peak in $d_{\rm cluster}$ is expected as the FoF center of mass of the identified galaxy clusters closely corresponds to the position of massive central galaxies that lie nearby, therefore leading to many galaxies sitting at these very low distance values. The 2nd peak is due to the satellite galaxies in more extended orbits around this central point. We confirm this in Fig. \ref{fig:SatCentDists} showing that the central galaxies lie at lower distances whilst the satellite galaxies extend to higher distances.

The distribution of $d_{\rm skel}$ in Fig. \ref{fig:FilClustDistributions} is uni-modal. As $d_{\rm skel}$ is based on the skeleton produced by {\tt T-ReX} rather than on the structure that is pre-defined within HR5 itself, it does not have a systematic population equivalent to central galaxies as $d_{\rm cluster}$. This means galaxies do not necessarily sit on the filament spines, like how they sit at the center of clusters; instead, they will be distributed around them. This uni-modality of $d_{\rm skel}$ is one of the expected results, as seen in \citet{Galarraga2020}, and we perform our analysis such that we match this literature. However, the distribution of $d_{\rm skel}$ has also been shown to appear as a bimodality in \citet{Song2021}, proposed to be due to statistical effects emerging from the skeleton computation.

\subsection{Cosmic Environments}
\label{sec:res_env}

We classify galaxies into three broad cosmic environments labelled as Nodes, Filaments and Voids based on $d_{\rm skel}$ and $d_{\rm cluster}$. In the following subsections, we discuss how we take specific cuts in $d_{\rm skel}$ and $d_{\rm cluster}$ to locate galaxies near large galaxy clusters, near filament cores, and far from any structure.

\subsubsection{Node population}
\label{sec:node}

To define the Node population we begin with the $d_{\rm cluster}$ dataset that includes galaxies within $2 \times R_{200}$ of halos with $M_{\rm tot}$ $\geq 10^{13} M_{\sun}$. This dataset already constitutes a viable representation of galaxies that belong to large clusters or nodes, however in Fig. \ref{fig:SatCentDists} we observe a significant scatter of up to $\sim 310$ ckpc between central galaxies and the FoF center of mass that $d_{\rm cluster}$ is measured relative to. This offset is shown in \citet{Gouin2021} to emerge from `unrelaxed' galaxy clusters that are in the process of merging and is commonly used as an indicator of the dynamical state of galaxy clusters (see also \citealt{Yoo2024} for more details). If two merging equal-mass clusters are considered one system by the FoF algorithm, the center of mass will be calculated to sit at a point between them both rather than near the position of either of the central galaxies associated with each cluster. This scenario leads to a larger $d_{\rm cluster}$ value, or an offset, for central galaxies. We aim for our node population to represent only galaxies within relaxed galaxy clusters that are virialized. We employ a similar cut to the one seen in \citet{Gouin2021} to achieve this. By normalizing the $d_{\rm cluster}$ values of the central galaxies to the $R_{200}$ value of the FoF halo they belong to, we achieve a measure of normalized offset, $\Delta_{r} = d_{\rm cluster}/R_{200}$, for each central galaxy. \citet{Gouin2021} takes a $\Delta_{r} < 0.07$ cut to represent the relaxed cluster situation. This work uses a slightly stricter cut of $\Delta_{r} < 0.05$ to define our relaxed clusters, \citep{Zhang2022}. With this cut employed, we now fully define our Node population as the galaxies that lie within $2 \times R_{200}$ of galaxy clusters with $M_{\rm tot}$ $\geq 10^{13} M_{\sun}$ where the central galaxy of the specific halo has $\Delta_{r} < 0.05$ from the FoF center. This analysis of $d_{\rm cluster}$ for the centrals leads to the conclusion that $d_{\rm cluster}$ values associated with `unrelaxed' clusters are not meaningful, as the FoF peak does not lie at the center of an individual cluster. Due to this, from here onwards, we limit $d_{\rm cluster}$ to only include galaxies that belong to the `relaxed' halos, with $\Delta_{r} < 0.05$. 

Using the calculated values of $\Delta_{r}$, we can also take a higher cut and create an environment representing galaxies in an `unrelaxed' halo environment. The different dynamical states of halos can lead to dramatic changes in the cluster environment, meaning the evolution of galaxies within will be affected. To create the `unrelaxed' population, just like for the `relaxed' clusters, we select galaxies that lie within $2 \times R_{200}$ of halos with $M_{\rm tot}\geq 10^{13} M_{\sun}$ and have $\Delta_{r} \geq 0.11$, corresponding to a large offset. This leaves us with the main Node population, comprised of galaxies within `relaxed' halos, and our `unrelaxed' population of galaxies. In Appendix B, Fig. \ref{fig:UnrelaxedvsRelaxed} shows two halos from the halo catalog that visually demonstrate our two cluster populations.

\subsubsection{Filament population}
\label{sec:fil}

We define our filament galaxy population based on $d_{\rm skel}$, the perpendicular distance from a galaxy to its nearest edge of the skeleton. The filament galaxies are then defined as galaxies with $d_{\rm skel}\leq 1 \, \text{cMpc}$. We choose this distance cut as inspired by other studies of filamentary structure in which typical filament radii lie in the range of $\sim 1 - 3 \text{cMpc}$, \citep{Wang2024,Galarraga2022}. As we aim to probe galaxies that lie in the cores of filaments, we choose a value of $d_{\rm skel}\leq 1 \, \text{cMpc}$ \citep{Galarraga2023}.
Filament lengths are also an important property to consider when discussing the large-scale structure across the cosmos, as this may also lead to slightly different environments for galaxies to evolve in. In {\tt T-ReX}, the length of each filament ($\ell_{\text{fil}}$) can be calculated by adding up the lengths of the individual edges in the skeleton that belong to each filament. We separate the filaments into short filaments ($\ell_{\rm fil} \leq 5$ cMpc) and long filaments ($\ell_{\rm fil} \geq 15$ cMpc) similar to the values taken by \cite{Galarraga2020}.  We compute each galaxy's perpendicular distance to its nearest long and short filament with these two filament datasets, giving us $d_{\rm skellong}$ and $d_{\rm skelshort}$, respectively.

\subsubsection{Void population}
\label{sec:void}

The void population comprises all galaxies at distances $d_{\rm skel} \geq 8 \, \text{cMpc}$ to the nearest filament.
The value of 8 cMpc is taken from the analysis of $d_{\rm skel}$ distribution in Fig. \ref{fig:FilClustDistributions} along with existing literature on the topic. \citet{Donnan2022} shows 10th and 90th percentiles in the distributions of $d_{\rm skel}$ and $d_{\rm cluster}$. By analysing {\tt IllustrisTNG}, the 90th percentile in their $d_{\rm skel}$ corresponds to a distance cut of ~$ \geq 6.39 \: \text{cMpc}$. Our distribution, as seen in Fig. \ref{fig:FilClustDistributions}, extends to slightly higher distances than seen in \citet{Donnan2022} such that we opt to use this higher value of 8 cMpc to suit our data better.

\subsubsection{Mapping the different cosmic environments}
\label{sec:map}

\begin{figure}
    \centering
    \includegraphics[scale=0.65]{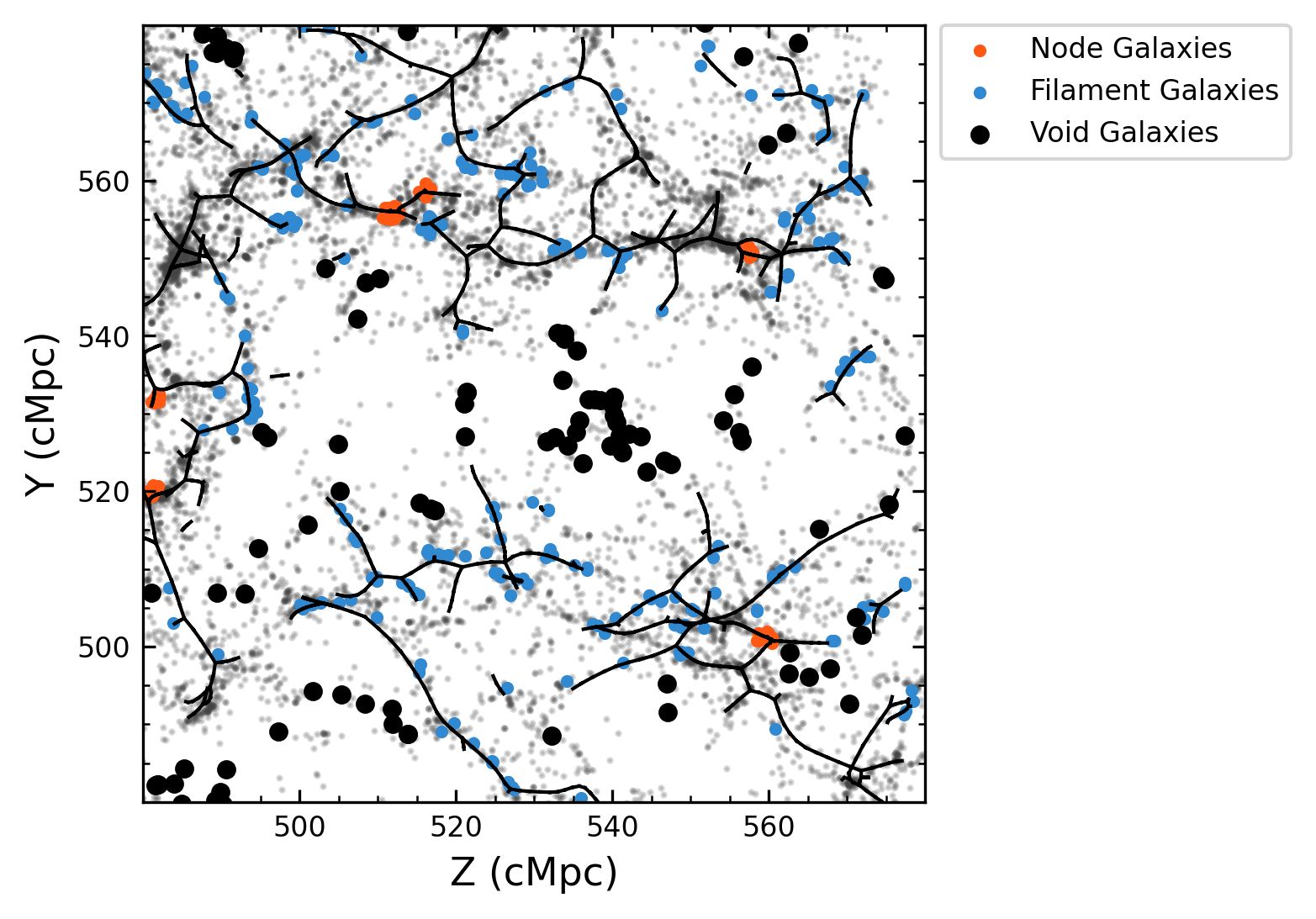}
    \caption{The three different types of environments displayed in a projected 20 cMpc slice of HR5. The galaxies have been coloured according to their corresponding environment (black corresponding to void galaxies, blue to filament galaxies, and orange to node galaxies). The figure shows that the galaxies have been correctly assigned to their respective environments corresponding to the global structure.}
    \label{fig:2dSliceEnvironment}
\end{figure}

With each of the three main populations defined, along with their sub-populations, Fig. \ref{fig:2dSliceEnvironment} shows how the different main galaxy populations are spatially distributed in a 2-D, $20$ cMpc slice of the simulation volume. Different colours correspond to different cosmic environments, with the thin, solid black lines representing the filaments in the skeleton, the black points show the position of void galaxies, the blue points show the position of filament galaxies, and the orange points show the position of the node, or cluster, galaxies.

\begin{figure}
    \centering
    \includegraphics[scale=0.65]{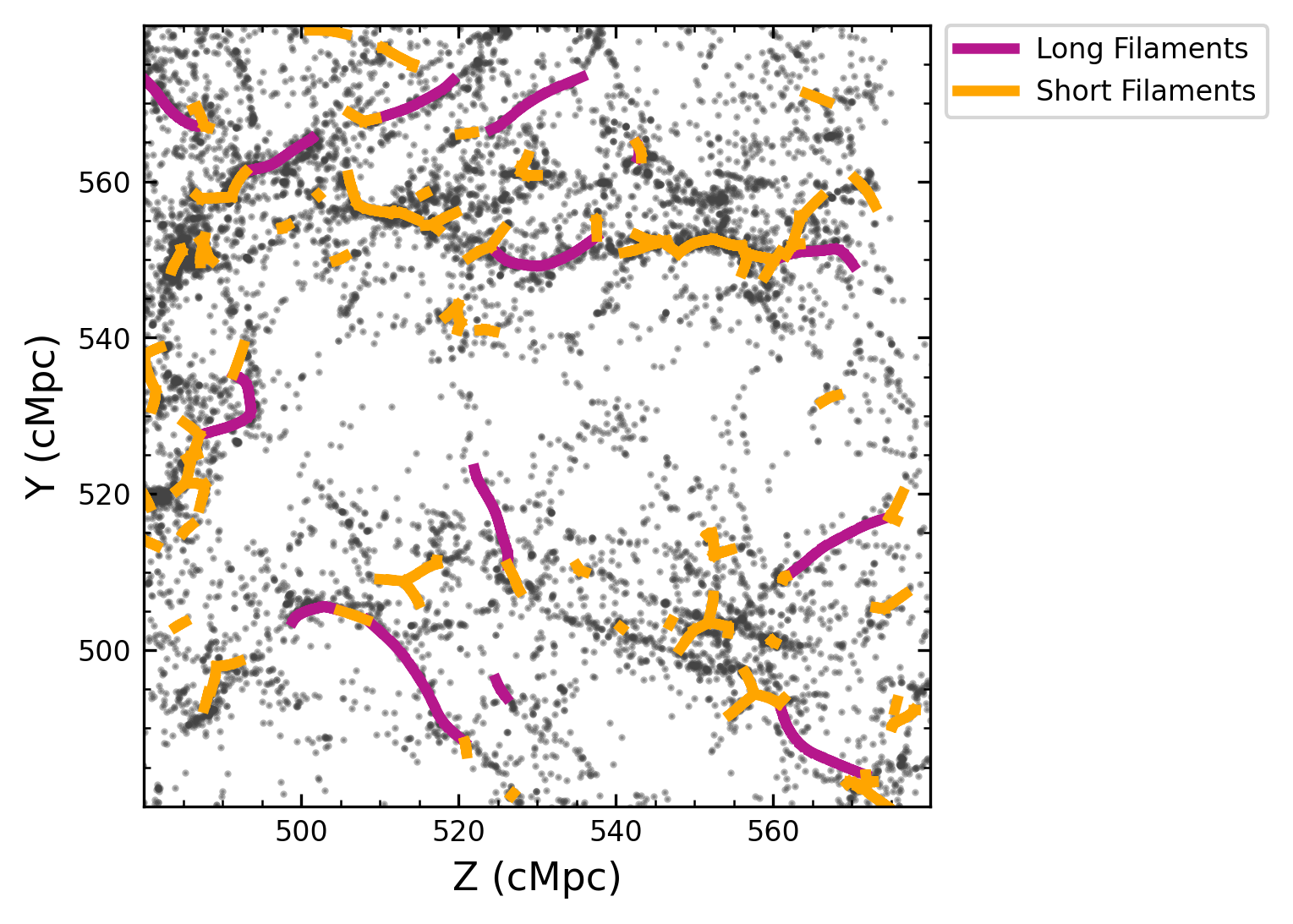}
    \caption{Distribution of long and short filaments in a projected 20 cMpc slice in HR5. The long filaments are shown in purple, representing filaments longer than 15 cMpc. The short filaments are shown in orange and are shorter than 5 cMpc.  Long filaments tend to form in the less dense regions, while short filaments reside in the more densely populated regions.}
    \label{fig:LongShortFilamentSlice}
\end{figure}

\begin{figure*}
    \centering
    \includegraphics[scale=0.65]{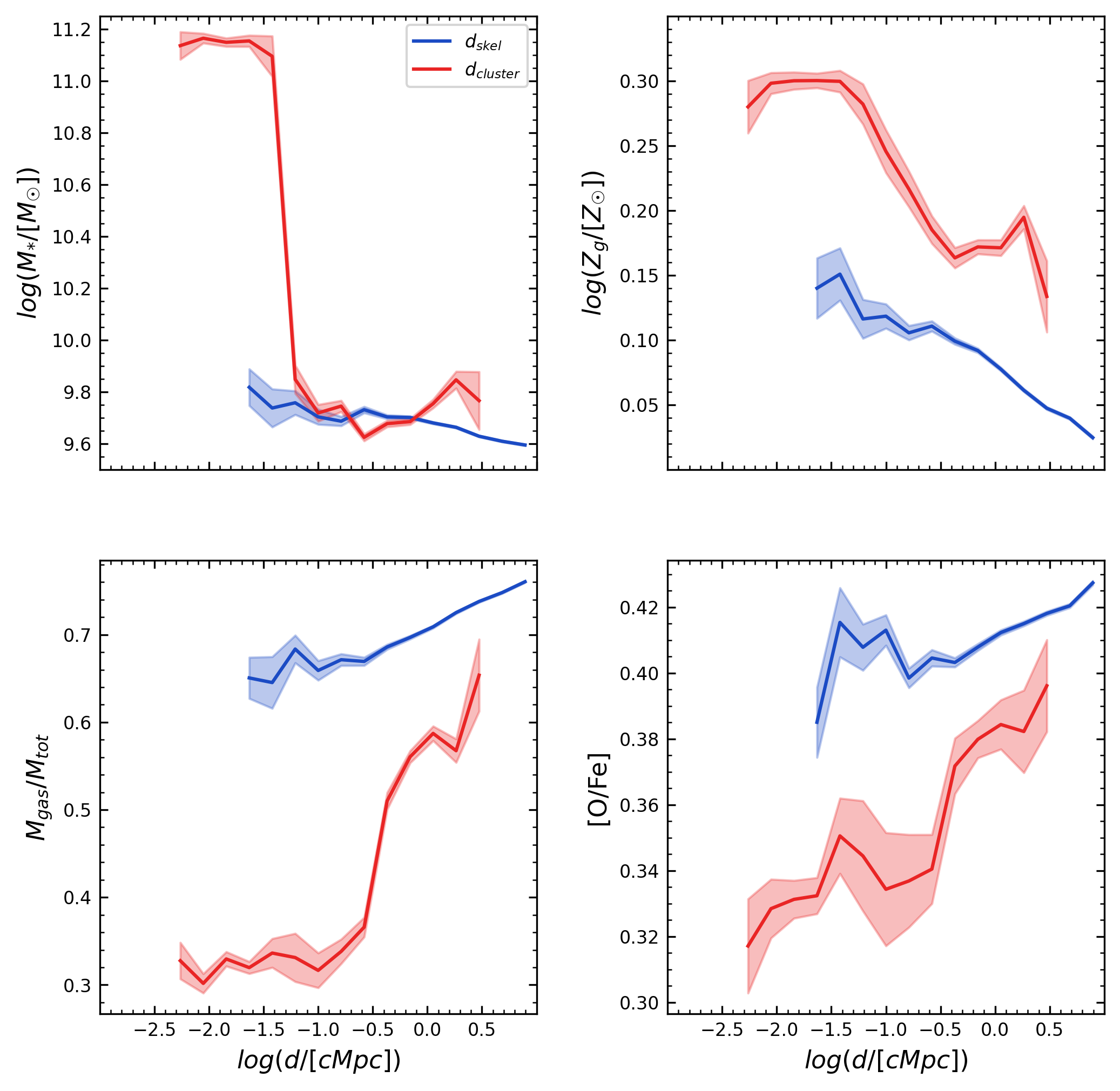}
    \caption{The figure shows how the median galaxy stellar mass (top left), gas metallicity (top right), gas fraction (bottom left) and [O/Fe] (bottom right) change as a function of the distance when considering different environments. 
    The blue curves show how the median values change in 20 equal bins as a function of $d_{\rm skel}$ when considering galaxies in the filament catalogue. The orange curves vary as a function of $d_{\rm cluster}$ for galaxies in the cluster catalogue. The shaded regions represent the standard errors on the median.}
    \label{fig:RadialPreRelations}
\end{figure*}

Fig. \ref{fig:LongShortFilamentSlice} highlights long and short filaments in a 20 cMpc slice of the simulation. Solid lines in purple and orange correspond to long and short filaments, respectively, whereas the grey points show the underlying galaxy distribution. The two figures qualitatively show that our methodology correctly identifies the different cosmic environments. Fig. \ref{fig:RadialPreRelations} shows how stellar mass, gas metallicity, baryonic gas fraction, hereby referred to as just gas fraction, and [O/Fe] vary as a function of $d_{\rm skel}$, in blue, and $d_{\rm cluster}$, in orange. Stellar mass and gas metallicity see negative correlations with both distance measures. However, the relationships observed in $d_{\rm cluster}$ are of a higher magnitude than seen with $d_{\rm skel}$. On the one hand, stellar mass shows an increase of $\sim1.6$ dex when comparing the satellite galaxies to the central galaxies, yet there is no conclusive linear trend when considering just the satellites. Gas metallicity shows an increase of $\sim0.16$ dex with a decreasing $d_{\rm cluster}$. They both show smaller increases of $\sim0.2$ and $\sim0.12$ dex with a decreasing $d_{\rm skel}$, with clear trends in both parameters. On the other hand, gas fraction and [O/Fe] show the direct opposite, namely a positive correlation with both distance measures. Gas fraction and [O/Fe] show decreases of $\sim 0.4$ and 0.08 dex, respectively, with a decreasing $d_{\rm cluster}$. Again, the magnitude of the trend reduces when considering $d_{\rm skel}$ with decreases of $\sim 0.1$ and 0.03 dex for gas fraction and [O/Fe], respectively. The observed relationships in the first two parameters and their magnitudes agree with results shown in \citet{Song2021} for stellar mass and \citet{Donnan2022} for gas metallicity.

\subsection{Environmental Dependence of the Mass-Metallicity Relation}
\label{sec:env_mzr}

\begin{figure}
    \centering
    \includegraphics[scale=0.65]{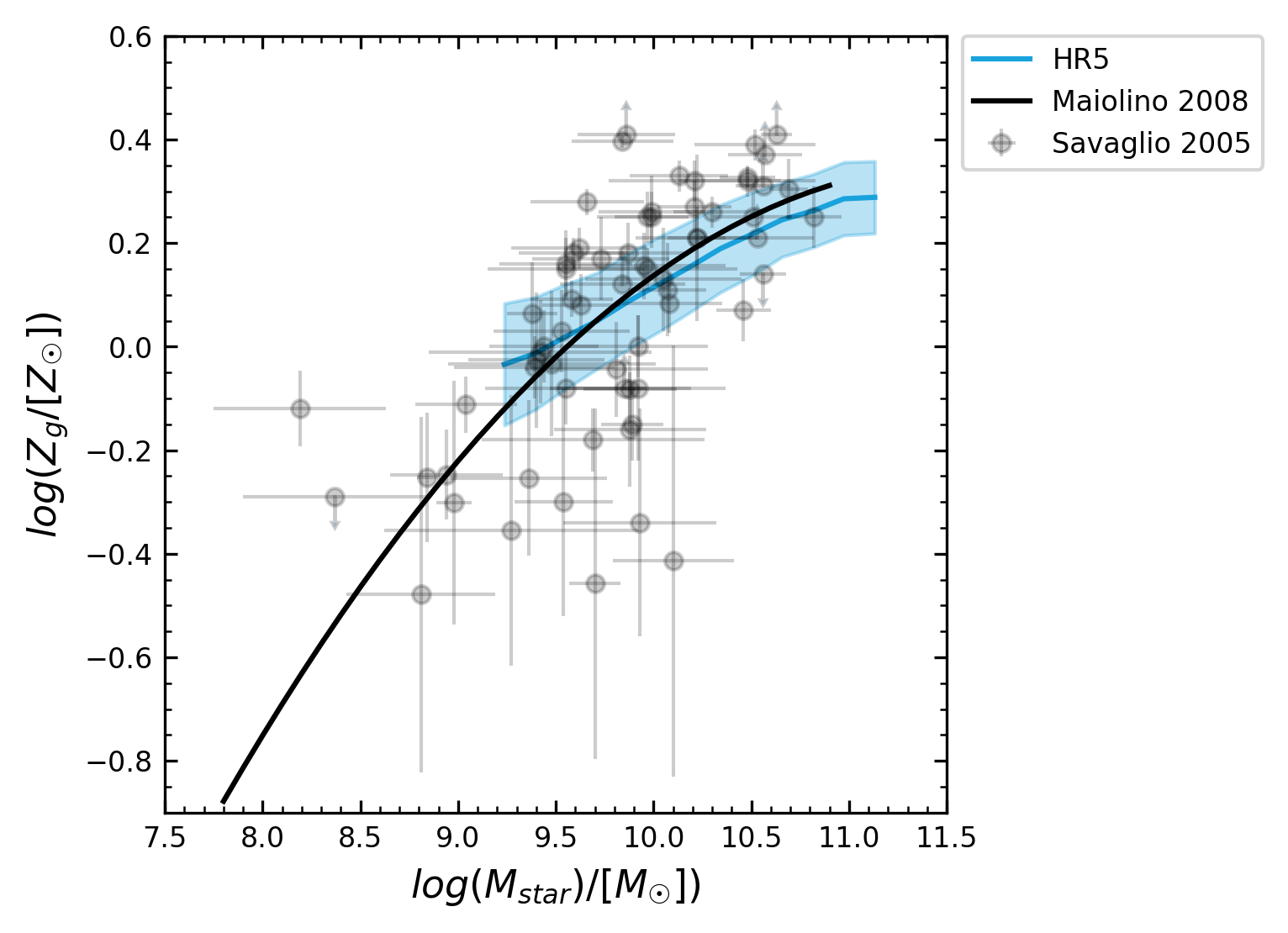}
    \caption{The MZR in HR5, in blue, with the shaded region representing a $1\sigma$ deviation around the median metallicity in each bin, overlayed with observational data. The black line shows the best-fitting model of \citet{2008Maiolino} to the observations of \citet{2005Savaglio} at redshift $z=0.7$ (grey points with error bars).}
    \label{fig:MZRComparison}
\end{figure}

We first present a short comparison of the MZR present in HR5 to observational data retrieved from \citet{2005Savaglio} and \citet{2008Maiolino} at a redshift of 0.7. It is shown in Fig. \ref{fig:MZRComparison} that HR5's MZR matches well with the average MZR reported by \citet{2008Maiolino} to fit the observational data of \citet{2005Savaglio}. Our analysis also indicates that the simulation predicts a similar scatter at high stellar masses to that seen in the data points of \citet{2005Savaglio} analysed by \citet{2008Maiolino}. This provides us confidence that our results are based on physics that is similar to that observed in the Universe itself, establishing a link to future observational studies. Note that in Fig. \ref{fig:MZRComparison} and throughout this work we assume the Solar abundances of \citet{asplund2009}.

\begin{figure}
    \centering
    \includegraphics[scale=0.65]{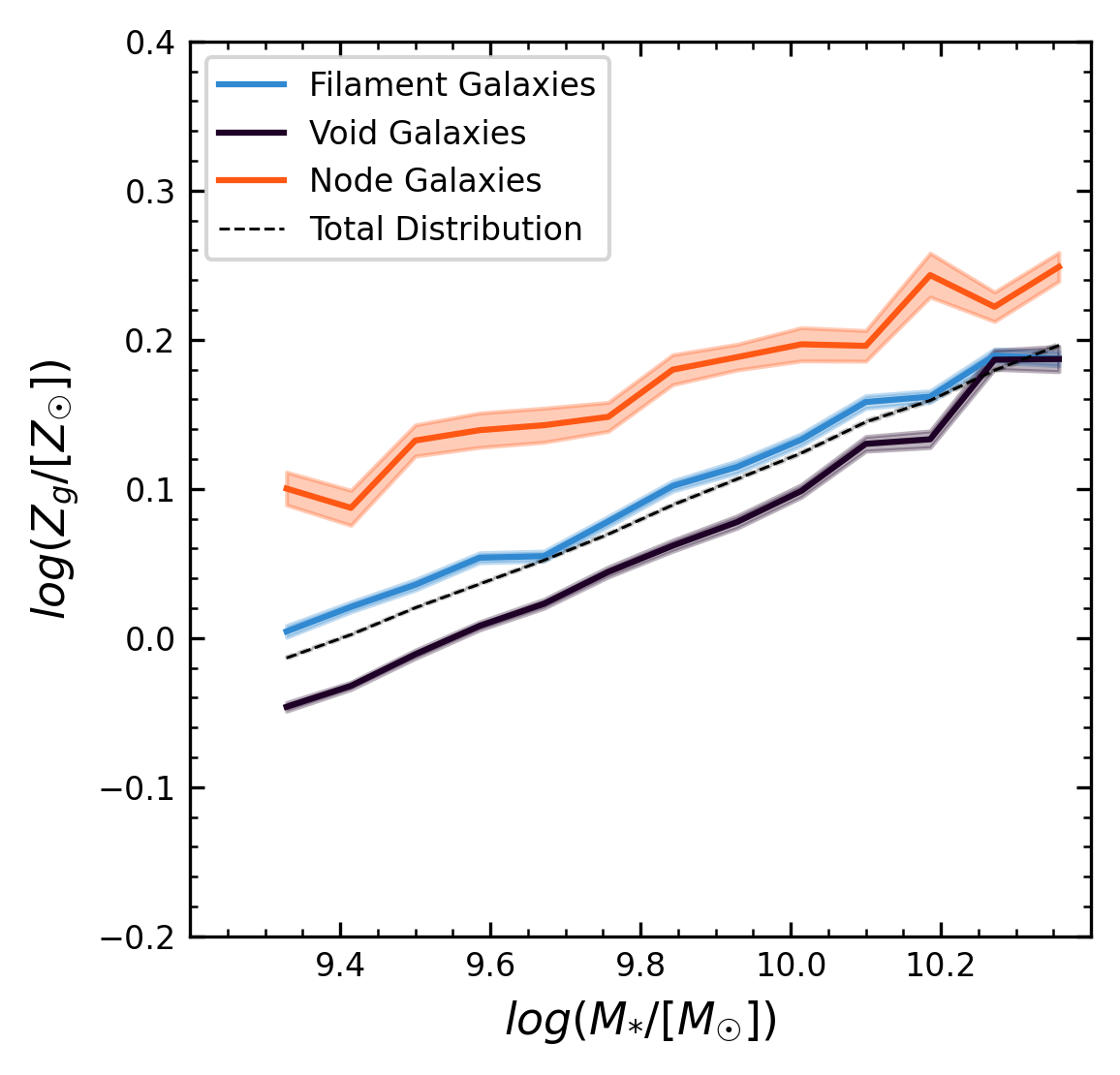}
    \caption{The galaxy stellar mass-gas phase metallicity relation in different environments, by splitting the sample into 15 equally-spaced, logarithmic bins in the galaxy stellar mass. The node population is shown in orange, the filament population in blue, and the void population in dark purple. The black dashed line shows the median MZR for the total galaxy distribution. The coloured shaded regions show the standard error on the median for each of the three environments.}
    \label{fig:MZRScatterNFV}
\end{figure}

The main section of our results begins by quantifying the effect of the three main cosmic environments (Nodes, Filaments and Voids) on the scatter of the median MZR in HR5. To perform this analysis, we calculate the median gas metallicity value of each cosmic environment within 15 equally-spaced, logarithmic stellar mass bins in the range $9.5 \leq \log(M_{\star}/\text{M}_{\sun}) \leq 10.4$ producing three MZRs. The results of our analysis are shown in Fig. \ref{fig:MZRScatterNFV}, indicating that the presence of different cosmic environments contributes significantly to the scatter seen in the total MZR. 

\begin{figure}
    \centering
    \includegraphics[scale=0.65]{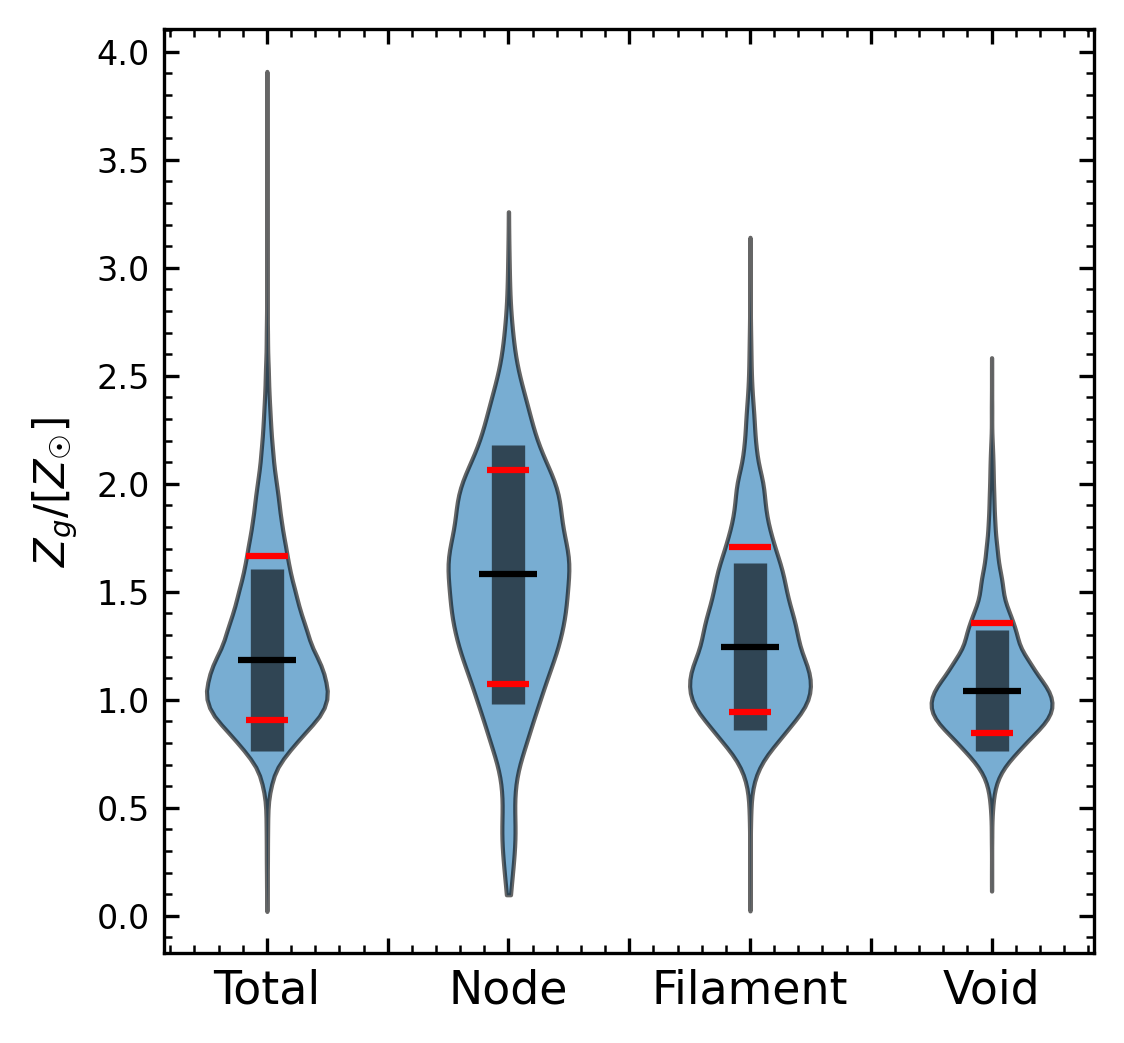}
    \caption{A violin plot displaying the distribution of the galaxy gas-phase metallicity, $Z_{g}$, associated with each environment. The thin horizontal black line shows the median of the distribution whilst the thick vertical black bar shows a 1$\sigma$ deviation. We also show percentiles of 15.9 and 84.1 in red to highlight the skew that exists for the Total, Filament, and Void populations when compared to the standard deviation.}
    \label{fig:stdenvironments}
\end{figure}

By averaging over the whole stellar mass range, the average relative standard deviation of the total MZR is $\langle \sigma_{Z} \rangle/\langle Z \rangle = 0.390$. Within each environment, the average relative standard deviations for their own MZRs are $0.596$, $0.363$, and $0.264$ for the Nodes, Filaments and Voids, respectively, which could be considered estimates for the intrinsic metallicity scatter. This relationship is displayed in Fig. \ref{fig:stdenvironments}, in which the progressive reduction in the standard deviation from the Node population to the Void population can be seen clearly. We also note that the Total, Filament and Void populations display a noticeable level of skew towards lower $Z_{g}$ values in the distribution, whilst the Node population is the least skewed.

The node population in Fig. \ref{fig:MZRScatterNFV} has the highest level of chemical enrichment for galaxies with stellar masses in the range $\log(M_{\star}/\text{M}_{\sun}) \lesssim 10.4$, with the median MZR sitting above that of the total galaxy distribution in the same stellar mass range. The filament population has intermediate median metallicities in the gas, showing slight enrichment above the total distribution for low stellar masses. For stellar masses in the range $\log(M_{\star}/\text{M}_{\sun}) \lesssim 10.3$, the void population has the lowest median gas metallicities among the considered cosmic environments. Interestingly, for $\log(M_{\star}/\text{M}_{\sun}) \gtrsim 10.2$, the Void and Filament populations are indistinguishable from each other, whilst the Node population remains at a higher enrichment, yet with less magnitude than at lower stellar masses.

The vertical scatter in the MZR due to the environment in Fig. \ref{fig:MZRScatterNFV} is more prominent at lower stellar masses. In particular, when moving from the median MZR of the node galaxies (orange line) to the MZR of void galaxies (dark purple line), the range of gas metallicities covers $\sim 0.14$ dex at $M_{\star} = 10^{9.25}\,\text{M}_{\sun}$, decreasing to $\sim 0.5$ dex at $M_{\star} = 10^{10.2}\,\text{M}_{\sun}$.

To quantify the environmental dependence of the MZR, we calculate how the residuals, $\delta Z_{g} = Z_{g} - \langle Z_{g} \rangle$, from the median MZR of the total galaxy distribution change as a function of stellar mass by considering galaxies in different environments. By averaging $\delta Z_{g}$ over the full stellar mass range, the node population has a median metallicity residual from the total MZR $\langle \delta Z \rangle_{node} = 8.4\times 10^{-2}$; the filament population is characterised by $\langle \delta Z \rangle_{fil} = 5.4\times 10^{-3}$, sitting a little closer to the median of the overall population, and the void population has $\langle \delta Z \rangle_{void} = -2.7\times 10^{-2}$. The Node population has the largest magnitude of deviation from the total MZR.

\begin{figure}
    \centering
    \includegraphics[scale=0.65]{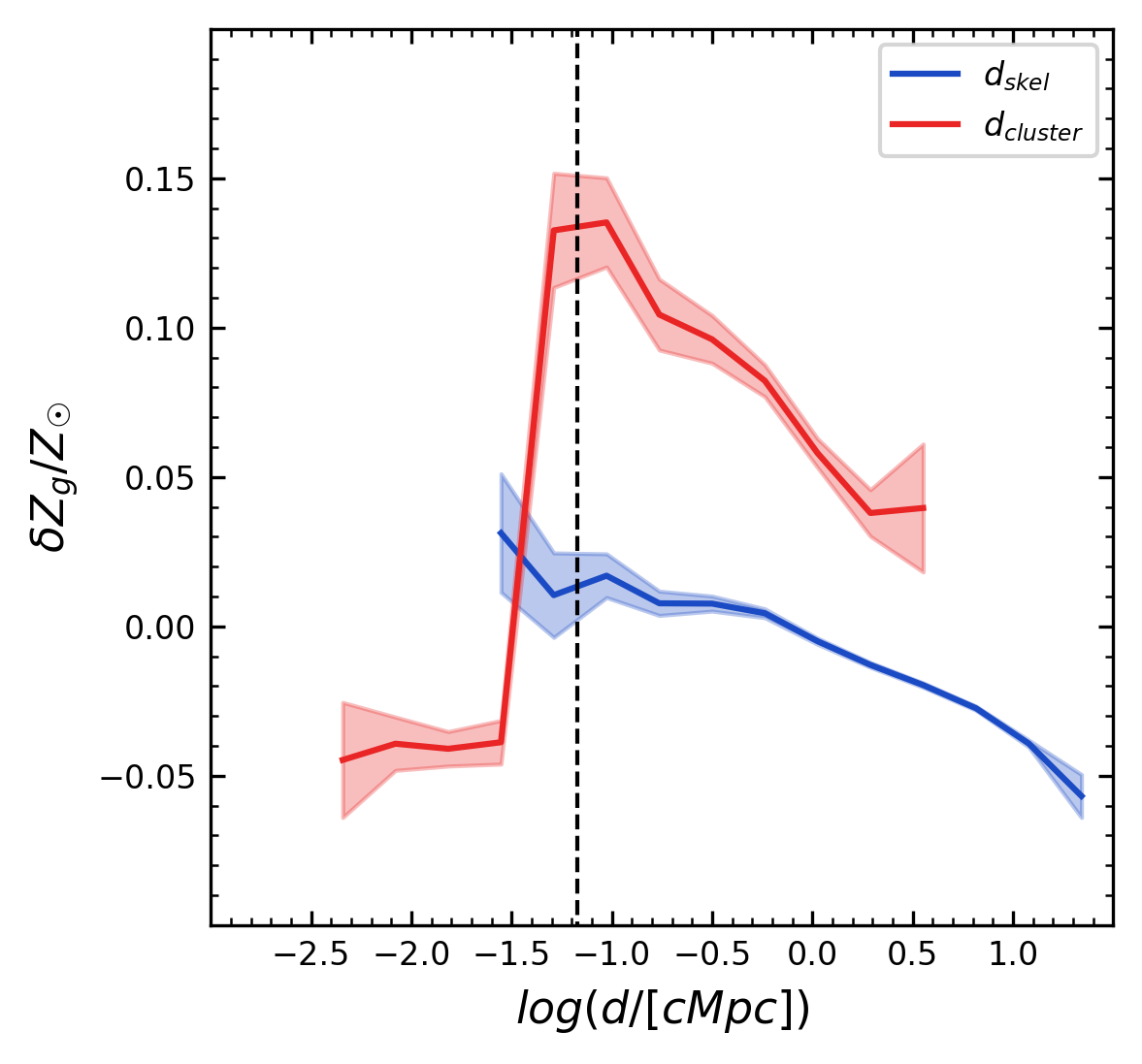}
    \caption{The figure shows how the gas metallicity residuals, $\delta Z_{\rm g}$, from a linear fit of the total MZR, $Z_{\text{fit}}(M_{{\star}}^{\mathcal{G}})$, change as a function of distance when considering different environments. $d_{\rm skel}$ represents each galaxy's distance to its nearest filament in the skeleton, shown in blue, and $d_{\rm cluster}$ represents each galaxy's distance to its nearest cluster with total mass larger than $10^{13}\,\text{M}_{\sun}$, shown in orange. The shaded regions represent the standard error on the median. Just as in Fig. \ref{fig:SatCentDists}, the vertical black dashed line denotes the distance at which central galaxies become the more prominent population relative to $d_{\rm cluster}$.}
    \label{fig:RadialDMZRClustFil}
\end{figure}

Fig. \ref{fig:RadialDMZRClustFil} quantifies how the residuals from the linear fit of the MZR of the total galaxy distribution, $\delta Z_{\rm g}$, change as a function of $d_{\rm skel}$ and $d_{\rm cluster}$, when considering filament and node populations, respectively. For this analysis, we first fit the total MZR by considering galaxies with stellar mass in the range $10^{9.2} \leq M_{\star} \leq 10^{10.4}\, \text{M}_{\sun}$ by using a linear regression model, $Z_{\text{fit}}(M_{\star})$. Then, for each galaxy $\mathcal{G}$ in the HR5 catalogue with stellar mass $M^{\mathcal{G}}_{\star}$ and gas metallicity $Z^{\mathcal{G}}_{g}$, the residual from the total MZR, $\delta Z_{g}$, is calculated as the difference between the galaxy gas metallicity and the metallicity, $Z_{\text{fit}}(M^{\mathcal{G}}_{\star})$. In Fig. \ref{fig:RadialDMZRClustFil}, we show how the median values of $\delta Z_{g}$ change when considering different bins of $d_{\rm skel}$ and $d_{\rm cluster}$.

$\delta Z_{g}$ from the total MZR in cluster galaxies decreases from $0.13$ to $0.05$ in the range $-1.3 \leq \log( d_{{\rm skel}/{\rm cluster}} / \text{cMpc} ) \leq 0.5$. This implies that as we approach the central regions of the cluster, galaxies become -- on average -- more metal-rich than the total galaxy population with the same stellar mass distribution at redshift $z=0.625$. When $\log( d_{{\rm cluster}} / \text{cMpc} )$ approaches $-1.3$, we observe that $\delta Z_{g}$ dramatically drops to negative values of $\sim  -0.04$. This distance value of $10^{-1.17}\,\text{cMpc}$ corresponds to the point in the $d_{\rm cluster}$ distribution at which the central galaxies become the prominent population (see Fig. \ref{fig:SatCentDists}). From Fig. \ref{fig:MZRScatterNFV} we can see that as stellar mass increases, the deviation from the MZR of the total population decreases. As the central galaxies are the highest stellar mass population, this large drop in $\delta Z_{g}$ at $10^{-1.3}\,\text{cMpc}$ in $d_{\rm cluster}$ is expected and matches what is predicted in the previous figure.
The deviation from the total MZR, $\delta Z_{g}$, in filament galaxies in Fig. \ref{fig:RadialDMZRClustFil} sees a slight positive deviation of $\sim 0.02 \, \text{dex}$ within $1 \, \text{cMpc}$ of filaments. With increasing $d_{\rm skel}$, $\delta Z_{g}$ becomes negative and continues falling to levels seen for the void population. Overall, this shows a slight metal enrichment in the cores of filaments, with less enrichment occurring at larger distances.
The void population, defined for  $d_{\rm skel} \geq 8\,\text{cMpc}$, shows the same low level of chemical enrichment that was observed in Fig. \ref{fig:MZRScatterNFV}.

In summary, proximity to clusters, $d_{\rm cluster}$, demonstrates a maximum $\delta Z_{g}$ of $\sim 0.13$ dex at a distance of $\log( d_{\rm cluster} / {\rm cMpc}) \approx -1.3$, and a minimum $\delta Z_{g}$ of $\sim -0.05$ dex for the central galaxies. Proximity to filaments, $d_{\rm skel}$, gives rise to a maximum $\delta Z_{g}$ of $\sim 0.02$ dex at small distances, only just breaking through to positive levels of enrichment in the context of the total MZR. At distances $d_{\rm skel}\geq 8 \text{cMpc}$, we see the largest negative $\delta Z_{g}$ of $\sim -0.5$ dex from the total MZR, corresponding to the void population.

\begin{figure}
    \centering
    \includegraphics[scale=0.65]{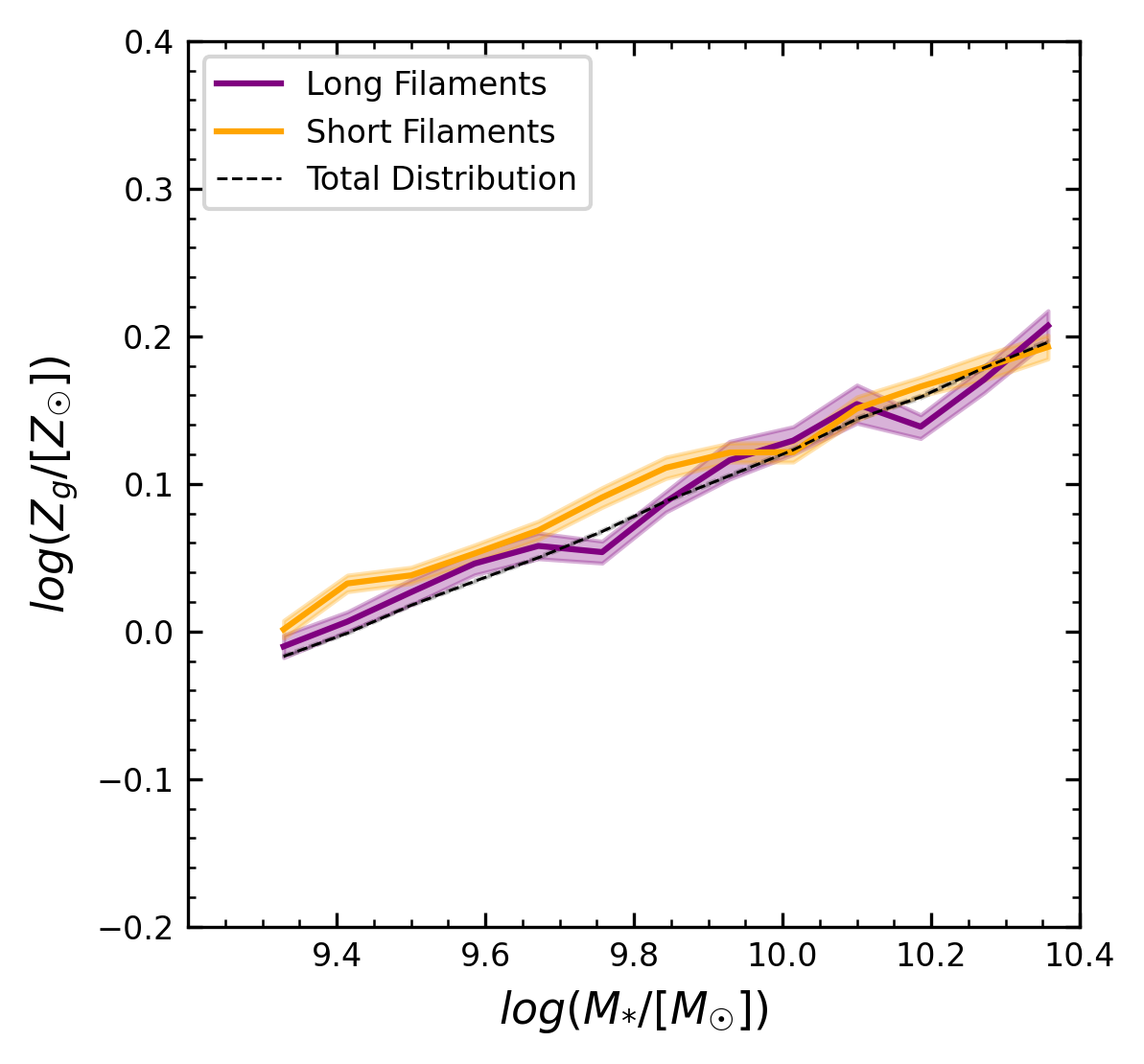}
    \caption{The median MZR for the long and short filament populations. The long filaments are shown in purple and correspond to galaxies that lie within 1 cMpc of filaments with a length longer than 15 Mpc. The short filaments are shown in orange, and conversely correspond to galaxies within 1 cMpc of filaments with a length shorter than 5 Mpc. The coloured shaded regions represent the standard error on the median. The median MZR for the total population of galaxies is shown as a black dashed line.}
    \label{fig:LongShortdMZR}
\end{figure}

In Fig. \ref{fig:LongShortdMZR}, we compare the median MZR of the long and short filament populations. Relative to the total MZR, short filaments show an average gas metallicity that is vaguely higher than the long filaments; however, both populations show a very slight metal enrichment compared to the total population of galaxies for $M_{\star} \leq 9.6$. Short filaments are typically found surrounding high-mass clusters and nodes of the cosmic web, explaining why we find marginally higher metallicities in short filaments than in the total population with a similar mass distribution; long filaments are more common in less dense regions (see Fig. \ref{fig:LongShortFilamentSlice}), which are instead characterised by lower average metallicities than the median MZR of the total population (see Fig. \ref{fig:LongShortdMZR}).

\begin{figure}
    \centering
    \includegraphics[scale=0.655]{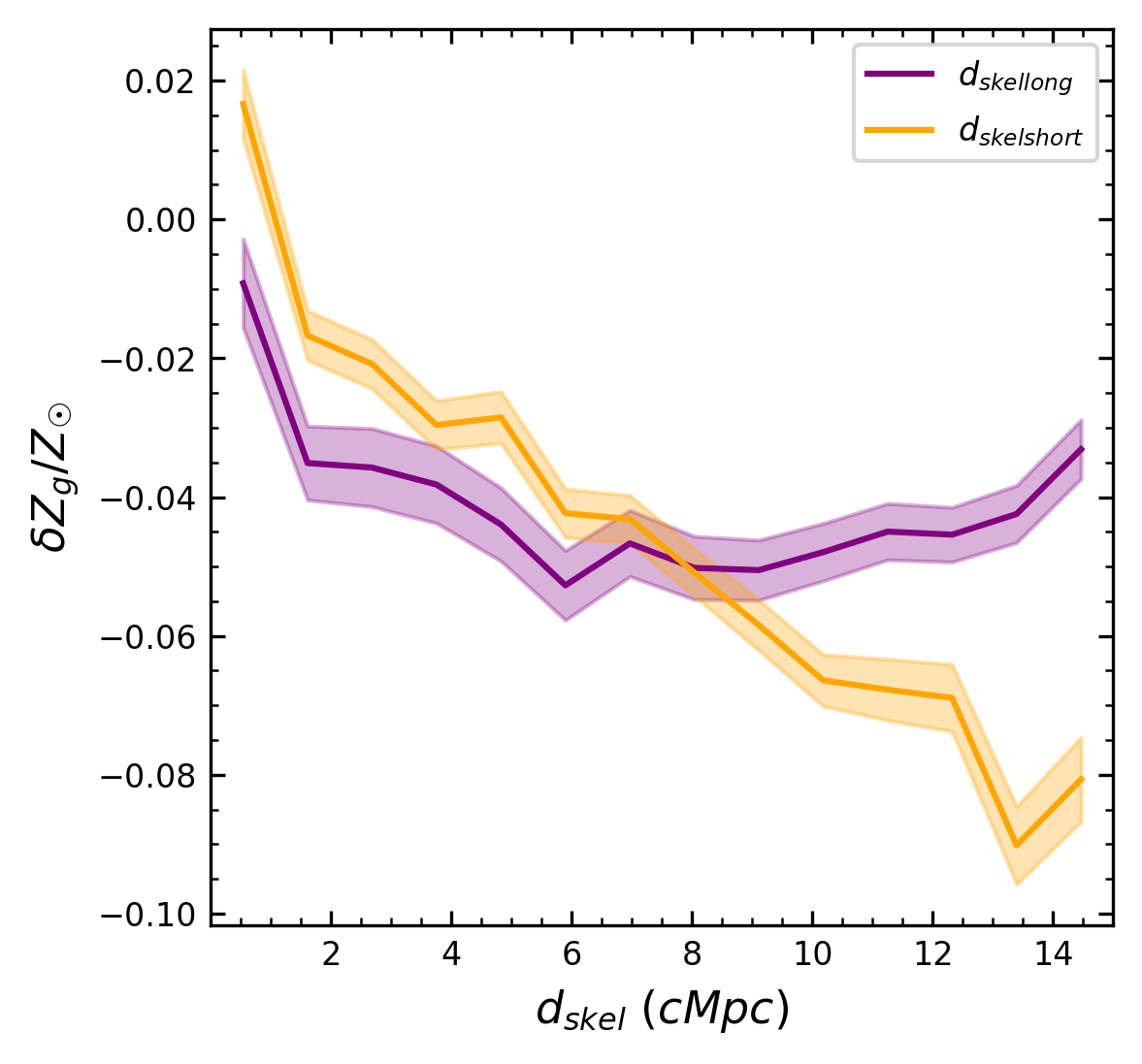}
    \caption{The median residual from the MZR, $\delta Z_{\rm g}$, for $d_{\rm skelshort}$ (in purple) and $d_{\rm skellong}$ (in orange). $d_{\rm skelshort}$ and $d_{\rm skellong}$ represent the perpendicular distance to a short or long filament of a galaxy, respectively. The shaded regions indicate the standard error on the median.}
    \label{fig:LongShortdMZRRadial}
\end{figure}

In Fig. \ref{fig:LongShortdMZRRadial}, we show how the deviation, $\delta Z_{g}$, in the gas metallicity of galaxies changes as a function of $d_{\rm skel}$, separating long filaments from short filaments in the analysis. Interestingly, the positive values of $\delta Z_{g}$ that are found for short filaments in Fig. \ref{fig:LongShortdMZR} only appear at distances $d_{\rm skelshort}\leq 1\,\text{cMpc}$ in Fig. \ref{fig:LongShortdMZRRadial}. Although the fraction of massive galaxies increases with a falling $d_{\rm skel}$, this result suggests that the majority of this trend emerges from the low stellar mass galaxies seen in \ref{fig:LongShortdMZR} where a positive residual is observed only for these galaxies.
The profiles of $\delta Z_{g}$ in long filaments show a much flatter trend. We still observe an increase in $\delta Z_{g}$ in the cores, at distances $\leq 1 \,\text{cMpc}$ from the cores of long filaments; however, for distances $ \geq 8$ cMpc, $\delta Z_{g}$ does not continue falling to values seen around short filaments and instead remains constant.

\begin{figure}
    \centering
    \includegraphics[scale=0.655]{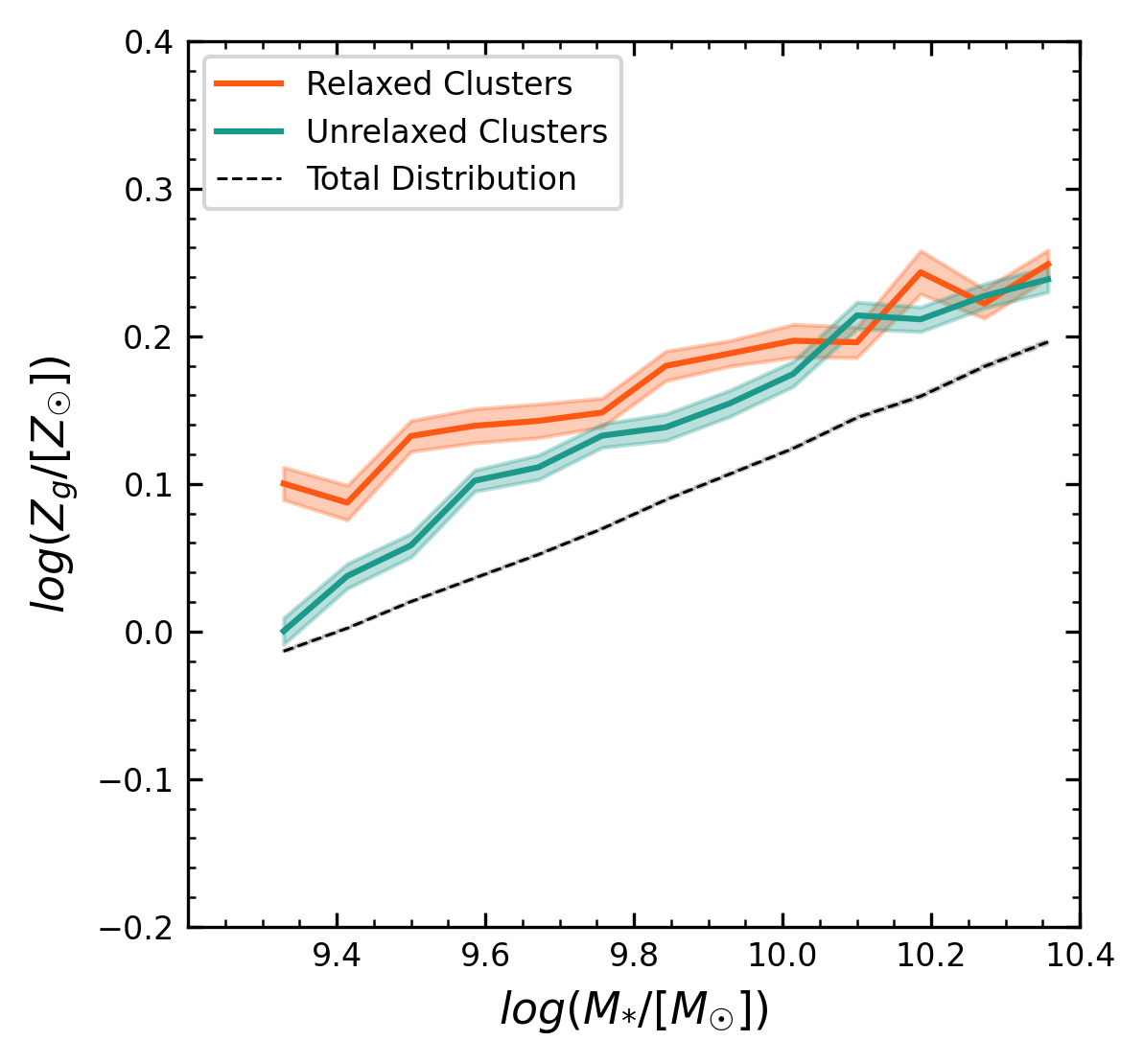}
    \caption{The MZR present in relaxed cluster environments, in orange, and unrelaxed cluster environments, in green. The black dashed line represents the MZR for the total population of galaxies irrelevant to their environment. The coloured regions around the orange and green lines represent the standard error on each, respectively. The black shaded region shows $0.5\space\sigma$ around the MZR for the total distribution, displaying the overall scatter in the relationship.}
    \label{fig:URClustMZRScatter}
\end{figure}

\begin{figure*}
    \centering
    \includegraphics[scale=0.52]{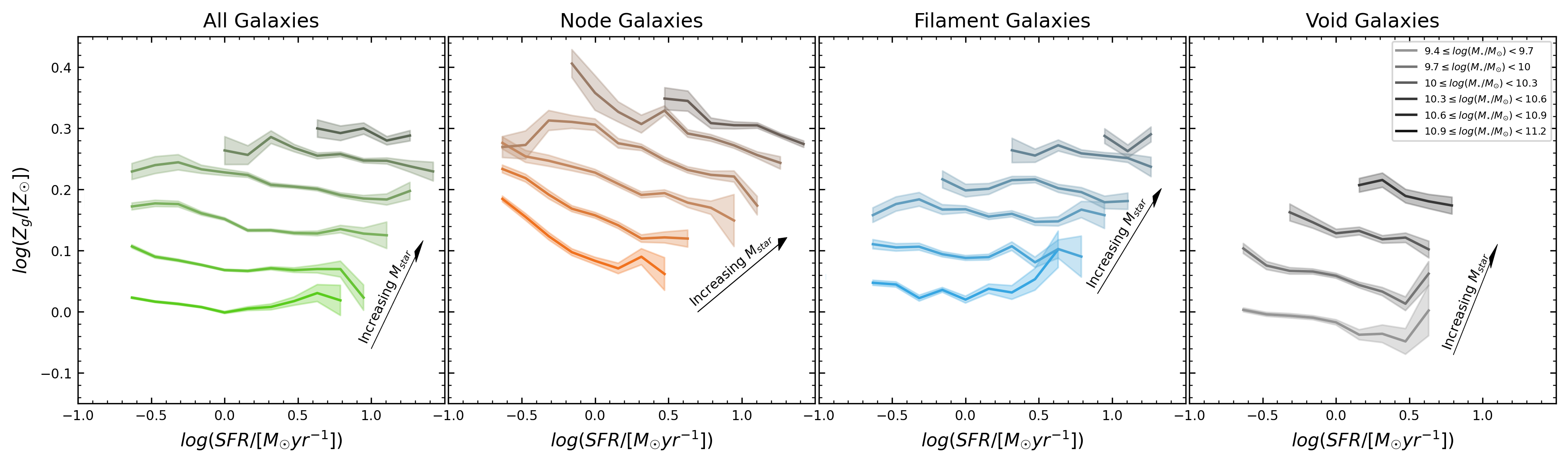}
    \caption{This figure shows how the SFR changes against the gas metallicity, $Z_{g}$, separating galaxies into six equally-sized logarithmic bins of $M_{\star}$ and different environments. From left to right, the first panel shows the total population of galaxies irrespective of environment (green lines), the second panel the node galaxies (orange lines), the third panel the filament galaxies (blue lines) and the fourth panel the void galaxies (black lines), demonstrating the predicted fundamental metallicity relation (FMR) in HR5 and the effect of cosmic environment on it. The shaded regions show the standard error on the median, where only bins that include more than 15 galaxies are shown. Darker colours indicate galaxies with larger stellar masses, as indicated in the legend. The arrow demonstrates the broad direction in the plot in which each consecutive line represents a higher bin of $M_{\star}$ than the last. The final panel only contains four lines as the higher stellar mass bins are not populated in the void regions.}
    \label{fig:FMR}
\end{figure*}

Looking to the effect on the MZR due to the differing dynamical state of clusters, Fig. \ref{fig:URClustMZRScatter} shows the main Node population (in orange), comprised of galaxies belonging to relaxed halos, compared to the galaxies belonging to unrelaxed halos (in green). It is clear that, for $M_{\star} \leq 10^{10}\,\text{M}_{\sun}$, galaxies belonging to relaxed clusters have higher average gas metallicity levels of enrichment, contributing to a positive scatter above the MZR for the total population across the full stellar mass range. For galaxies belonging to unrelaxed clusters, a slight difference emerges, with significantly less enrichment when compared to the relaxed population; however, the relaxed population continues to have positive enrichment above the total MZR over the whole stellar mass range. At $M_{\star} = 10^{9.35}\,\text{M}_{\sun}$ the dynamical state of the cluster accounts for $\sim 0.1 \,\text{dex}$ scatter in the total MZR, quickly falling to smaller deviations for higher stellar masses. For $M_{\star} \geq 10^{10}\,\text{M}_{\sun}$, the two populations are indistinguishable. Again, it is clear from this figure that the effect is far more prominent for low stellar mass galaxies. 

\subsection{Environmental Dependence of the Fundamental Metallicity Relation}

When discussing galaxy evolution, and more specifically metallicity, it is important to also consider the SFR of galaxies as a clear link between the two properties has been established in many previous studies \citep{Mannucci2010,Cresci2012,Yates2012,HaydenPawson2022}. These works also show that a fundamental metallicity relation (FMR) exists between the three properties, stellar mass, gas metallicity and SFR. To ensure the completeness of our work in this paper, we also present an analysis of the FMR that exists in HR5, and assess how our cosmic environments effect the relation. The first panel on the left in Fig. \ref{fig:FMR} shows the FMR in HR5 for our total galaxy population irrelevant of environment. Within this plot a weak negative correlation exists for the intermediate mass bins, with the gas metallicity, $Z_{g}$, changing by $\approx 0.05\,\text{dex}$ for SFRs in the range $-0.5 \lesssim \log(\text{SFR}/\text{M}_{\sun}\,\text{yr}^{-1}) \lesssim 0.2$, however the relation becomes increasingly weaker in the lowest and highest mass bins. From left to right, the three other panels show the predicted FMR for galaxies associated with nodes, filaments and the voids respectively. When considering galaxies in nodes, we observe the trends between $Z_{g}$ and SFR to strengthen, showing a $\approx 0.2$ dex decrease in $Z_{g}$ for the $9.4 \leq \log(M_{\star}/\text{M}_{\sun}) < 9.7$ mass bin and SFRs in the range  $-0.5 \lesssim \log(\text{SFR}/\text{M}_{\sun}\,\text{yr}^{-1}) \lesssim 0.2$. This reflects something more similar to the trends observed in \citet{Mannucci2010}. For galaxies in filaments, the trends in the simulated galaxies flatten, observing no correlation between $Z_{g}$ and SFR for a given stellar mass. Finally, when we look to the galaxies in voids, the trends with SFR can be seen, however they are weak, with $\approx 0.03 - 0.06$ dex changes in $Z_{g}$ at the given stellar masses.

Our analysis indicates that cosmic environment does have an effect on the tilt of the 3D plane where galaxies sit according to their $Z_{g}$, $M_{\star}$ and SFR drawn by the FMR. When considering galaxies of fixed $M_{\star}$, while $Z_{g}$ has a weak correlation with the SFR within filaments and voids, $Z_{g}$ has a stronger dependence on the SFR in nodes; this causes the FMR plane of node galaxies to have a larger tilt angle. On the whole our results suggest that in Nodes, SFR is driving some of the scatter observed in the MZR, however in Filaments and Voids, this is not particularly the case, with the environment itself setting the level of chemical enrichment. For a given $M_{\star}$, the ISM of galaxies within more virialized nodes has been affected more by environmental processes and their associated feedback in the past, increasing the predicted $Z_{g}$ and giving rise to lower gas fractions and SFRs, which explains the trends seen in Fig. \ref{fig:FMR} for the node galaxies.

\section{Discussion}
\label{sec:disc}

\subsection{Radial relations with Gas Fraction and [O/Fe]}

From Fig. \ref{fig:RadialPreRelations}, proximity to the cores of filaments and the centers of relaxed clusters clearly affects galaxy properties, including baryonic gas fraction and [O/Fe]. When considering the gas fraction of galaxies, we see a large reduction of $\sim 0.4$ with $d_{\rm cluster}$ and a smaller reduction of $\sim 0.1$ with $d_{\rm skel}$. As expected, proximity to clusters, as they are one of the extreme environments in the web with a very large density contrast to the surrounding field, leads to a much larger magnitude of change than seen for filaments. The gas fraction of galaxies represents how much of the baryonic budget is in the form of gas, calculated as $f_{\rm gas} = M_{\rm gas}/(M_{\rm gas}+M_{\rm star})$. For the lowest values of $d_{\rm cluster}$, we are probing the central galaxies of the nodes/clusters. These central galaxies are older, have higher metallicity and higher stellar mass, and have a stronger burst of star formation in the far past. Over their longer lifetimes, feedback from AGN and SN explosions in the simulation has effectively removed the gas from the system, leading to lower gas fractions and higher metal content. AGN feedback is the more dominant scheme in this case of central galaxies in HR5, as SN feedback is mostly effective in smaller halos with shallow potential wells \citep{Somerville2008,Dubois2016}. These smaller halos are typically satellite galaxies. Here, AGN plays less of a role in removing gas with ram pressure stripping, and SN explosions are becoming the main mechanisms that do so.

Our analysis of the filament environment in the HR5 simulation shows that the filament population has significantly higher gas fractions at all values of $d_{\rm skel}$ but still reports lower gas fractions with higher proximity to the filament's core. This trend can be qualitatively explained by the same as described previously for clusters: galaxies with higher average stellar mass exist more frequently in the cores of filaments, where the local density of galaxy tracers is higher than in the field. Higher galaxy densities lead to more interaction between the galaxy and the environment, and this leads to increases in SFR, consuming gas in the reservoir more efficiently and resulting in more SN events, pushing gas out of the galaxy and increasing the average metal content in the gas, \citep{Mahajan2018, Liao2019, 2020Singh}.

When looking at the trend in the average gas-phase [O/Fe] ratios we again see positive correlations with $d_{\rm skel}$ and $d_{\rm cluster}$. Cluster galaxies show a reduction in [O/Fe] with decreasing $d_{\rm cluster}$ of $\sim 0.08 \, \text{dex}$ whereas the filament population shows a weaker trend with a reduction of $\sim 0.03 \, \text{dex}$. [O/Fe] is a chemical clock typically used as a proxy for the age of a galaxy in observations \citep{Matteucci1986, Miglio2021}. Although we have access to the true age of these galaxies in the simulation, we use [O/Fe] to provide a reasonable comparison to observational work. Lower [O/Fe] gas values in the centers of clusters confirm that this is an older population of galaxies; with increasing $d_{\rm cluster}$, we see [O/Fe] rise, meaning the galaxies are getting younger on average as we move away from the cluster center. Due to these central galaxies being older, they have had more Type Ia SNe occur, increasing their iron content and reducing [O/Fe]. This is also the case in the filament population, with filament cores showing a smaller magnitude reduction in [O/Fe]. Comparing the cluster population to the filament population shows us that filament galaxies are a younger population of galaxies, with less contribution from Type Ia SNe. This illustrates the hierarchical nature of the large-scale structure and galaxy evolution. The overall radial gradient that exists, and the fact that it is stronger for clusters than filaments, shows that the density contrast to the field is higher in clusters, in agreement with previous studies, \citep{Cautun2014}. It also shows that proximity to the centers of clusters and the cores of filaments in the simulation exhibits a trend with the average age of the dominant galaxy stellar populations, as an increase in the average level of SFR in the galaxy causes a more prominent contribution from Type Ia SNe on long timescales, which could possibly help to further quench the star formation activity which would enhance the iron content of the galaxies that reside there.

\subsection{Scatter in the Mass-Metallicity Relation}
\label{sec:scat}

The measured cosmic environments noticeably affect the vertical metallicity scatter in the predicted MZR at $ z = 0.625$ in HR5. At low stellar masses, in the range $9.3 \leq \log(M_{\star}/\text{M}_{\sun})\leq 9.6$, the total vertical deviation from the median metallicity in the Void population to the median metallicity in the Node population is a $\sim 0.45 \sigma$ or, equivalently, by $0.14 \:\text{dex}$ contribution to the overall metallicity scatter in the MZR. At higher stellar masses, in the range $10.1 \leq \log(M_{\star}/\text{M}_{\sun})\leq 10.4$, the metallicity deviation from the total MZR is reduced, and the levels of chemical enrichment are more comparable to the total distribution of galaxies. Our result that lower stellar mass galaxies contribute more to the scatter seen in the MZR suggests that low-mass galaxies are more sensitive to environmental effects than their high-mass counterparts, agreeing with previous studies (e.g., see \citealt{Mouhcine2007, Petropoulou2012}). This low stellar mass dependency could be depend on the link explored in \citealt{Yang2024} building on the key relationship between gas-fraction and gas-metallicity. It is observed, using galaxies from SDSS DR8, that scatter in the MZR strongly depends on gas-mass for low stellar masses with the trend vanishing at $M_{\star} = 10^{10.5}$. As our cosmic environments are inherently linked to different levels of gas-accretion, local densities and galaxy interactions, it is reasonable to expect to see a similar relationship with stellar mass.

\subsubsection{Contribution from the node population}
\label{sec:con_node}

Galaxies that reside within nodes or clusters in HR5 exhibit higher levels of chemical enrichment compared to galaxies within filaments and voids (see Figs. \ref{fig:MZRScatterNFV} and \ref{fig:RadialDMZRClustFil}). 
The nodes of the cosmic web are very dense regions of galaxies, where there is a higher frequency of events like galaxy perturbances, galaxy mergers \citep{Toomre1972,Schwiezer1982,LHullier2012}, tidal and ram pressure stripping \citep{Jhee2022,2019Singh,2024Singh,GunnGott1972}, some of which can increase the SFR and in turn the average gas metallicity \citep{Torrey2019}. This higher gas metallicity value for node galaxies may also be a product of the deeper gravitational potential they exist within. A deeper potential can make it harder for galaxies to lose their high-metallicity gas and may also interrupt the accretion from low-metallicity gas from outside the halo \citep{Dekel2009, 2014Peng} leading to higher metallicities. \citealt{Wang2023} reported that massive galaxies, in massive halos, are more metal-poor at a fixed stellar mass, whilst low-mass galaxies in these massive halos are more metal-rich, reflecting what we see in \ref{fig:MZRScatterNFV}. Our results also agree with the findings of \citet{Donnan2022}, who compared the average gas metallicity of galaxies within different environments both in the SDSS observations and in the IllustrisTNG simulation, determining higher average gas metallicities within node galaxies. 

When considering the dynamical state of clusters, Fig. \ref{fig:URClustMZRScatter} showed that at low stellar masses, $M_{\star} = 10^{9.35}\,\text{M}_{\sun}$, these environments contributed a total of $\sim \space 0.1 \text{dex}$ to the scatter in the MZR. This deviation between galaxies in unrelaxed and relaxed clusters quickly dropped to $\sim \space 0.05 \, \text{dex}$ for higher stellar masses in the range $10^{9.6} \leq M_{\star} \leq 10^{10.1}\,\text{M}_{\sun}$. For the highest stellar masses, the two environments are shown to be indistinguishable from each other.
\citet{Soares2019} finds that, on average, galaxies that belong to unrelaxed clusters, more specifically, the secondary halo in the system, demonstrate a younger stellar age than galaxies in relaxed clusters. This, therefore, means that these galaxies have had less time to undergo processes that increase metallicity, leading to a less metal-enriched population that we see in Fig. \ref{fig:URClustMZRScatter}.

On the other hand, relaxed clusters are formed from an older population of galaxies and are seen to exist in a slow accretion phase. This large amount of available relaxation time has allowed the metal enrichment to occur, \citep{Gouin2021}. It is also expected that these relaxed clusters will have higher local densities and, as such, an increased metallicity from these galaxy interactions.

\subsubsection{Contribution from the void population}
\label{sec:con_void}

The void population in our analysis demonstrates lower chemical enrichment. Previous works have not been able to establish a consensus on the effect of large-scale environments on the average gas metallicity of void galaxies. \citet{Kreckel2015} and \citet{Wegner2019} found void galaxies to have average gas metalicities and SFR properties, respectively, that are consistent with samples of galaxies of similar stellar mass within more crowded environments, whilst \citet{Pustilnik2011} found void galaxies to exhibit, on average, $30 \%$ lower gas metallicity values than galaxies in more dense environments with similar stellar mass, which aligns with our results. Our results also qualitatively agree with the measured stellar metallicities of \citet{Dominguez2023} in a sample of void galaxies in the Calar Alto Void Integral-field Treasury surveY (CAVITY) at redshifts $0.01 < z < 0.05$, as well as with the measured gas metallicities of \citet{Donnan2022} for void galaxies in SDSS at redshift $z=0.071$. \citet{Donnan2022} also found similar results in analysing the MZR of void galaxies in the TNG simulation. 

As void regions are the direct opposite of nodes within the context of local density \citep{Shim2023}, we expect to see far less frequent interaction between galaxies in this environment \citep{2012Jian}. Due to fewer interactions with other (sub)structures in the environment, one would expect less SFR enhancement, leading to less chemically enriched gas released by SNe and stellar winds into the ISM, and hence lower average gas metallicity values within the galaxy compared to denser environments.

It has also been proposed that two modes of gas accretion (`cold' and `hot' accretion) could be at play in nature \citep{Keres2005}. In this scenario, low-mass galaxies in less dense regions typically experience cold accretion, while high-mass galaxies in clusters would be dominated by hot accretion. This cold accretion would imply that low-mass galaxies in voids accrete pristine gas through smaller filamentary structures, sustaining their SFR on longer timescales than galaxies in nodes and filaments but at a lower pace, giving rise to lower average gas metallicities (see also \citealt{McQuinn2015a, McQuinn2015b} for an observational perspective).

\subsubsection{Contribution from the filament population}
\label{sec:con_fil}

Galaxies in the filament environment (with distance to the nearest filament in the skeleton, $d_{\rm skel}, \leq 1 \text{cMpc}$) show slightly higher metal enrichment for low stellar masses above the MZR seen for the total distribution of galaxies (see Figs. \ref{fig:MZRScatterNFV} and \ref{fig:RadialDMZRClustFil}). Several studies have recently compared the average metallicity of filament galaxies to their counterparts in the void, also analysing how the galaxy metallicity changes as a function of filament proximity, similarly to our Fig. \ref{fig:RadialPreRelations}. For example, \citet{Bulichi2023} concluded that, compared to the field, filament galaxies are more metal-rich, likely due to their higher-density environment. 

\citet{Donnan2022} found that filament proximity has little-to-no environmental effect on the predicted metallicity of galaxies in the MZR at $z=0.1$ within the IllustrisTNG simulation, yet the analysis of SDSS galaxy gas metallicities at $z=0.071$ showed a more noticeable change when considering different environments.

As discussed in \citet{Castignani2022}, filaments represent an intermediary environment between voids and nodes, with local densities of galaxies sitting between those of the other two environments yet slightly higher than the median of the total distribution within the core; this gives rise to a lower frequency of galaxy interactions than in environments closer to nodes, yet higher than seen in the voids and as such slightly increased enrichment. Overall, the filament population itself contributes very little to the scatter in the MZR, $\sim 0.01-0.02 \text{ dex}$ enrichment for low stellar masses, but as mentioned, since this is an intermediary environment, this is expected.

\subsubsection{The metallicity of short vs. long filaments}
\label{sec:shortlong}

The length distribution of filaments within different environments and how this length value correlates with galaxies' stellar and gas properties are currently the subject of intense investigation. \citet{Galarraga2021} using IllustrisTNG at $z=0$ studied how the filament length correlates with gas properties, finding that shorter filaments exhibit higher temperatures and pressures than longer filaments. Observationally, \citet{Castignani2022} showed that long filaments with $\ell_{\rm fil}> 17 h^{-1}\,\text{Mpc}$ tend to be relatively thin in radius ($R_{\rm fil}< 1 h^{-1} \text{Mpc}$) with a low-density contrast relative to the surrounding field. In Fig. \ref{fig:LongShortFilamentSlice}, we clearly see that the short filaments reside in the more densely populated regions of the web, near nodes, whilst the long filaments are found further out in the field in low-density regions. 

By introducing a length cut to distinguish between short and long filaments, we see marginal deviations between the two populations emerge in the mass range $9.3 \leq \log(M_{\star}/\text{M}_{\sun})\leq 10$ for short filaments, and across the full stellar mass range for long filaments. The short filament (with $\ell_{\text{fil}} \leq 5 \,\text{cMpc}$) population shows slightly higher enrichment than the long filament population ($\ell_{\text{fil}} \geq 15 \,\text{cMpc}$), with both populations above the MZR for the total distribution as seen in Fig. \ref{fig:LongShortdMZR}.

On the one hand, it is very possible that filament length, in this context, is acting as a proxy for the cosmic environment or, more specifically, as a proxy for $d_{\rm cluster}$, as short filaments are typically found connecting large clusters of galaxies, at higher proximity to nodes, \citep{Galarraga2020,Galarraga2023}.
Therefore, galaxies associated with short filaments will be subject to a higher frequency of galaxy-galaxy interactions, leading to higher levels of chemical enrichment. Instead, galaxies near long filaments will experience the opposite due to their typically less dense environment.

On the other hand, another proposed explanation is that the short filaments are thicker, more established and have a higher gravitational potential than the long filaments \citep{Galarraga2022}. This means that rather than the ambient cosmic environment being the cause of higher chemical enrichment, the environment of the short filament cores themselves is significantly different from the larger region they exist within. Our result in Fig. \ref{fig:LongShortdMZRRadial} supports this scenario. We show that $\delta Z_{g}$ in short filaments increases by $\sim 0.12 \:\text{dex}$ when moving from $d_{\rm skel}\sim 14\,\text{cMpc}$ to high proximity to the filament core, suggesting that it is the proximity to the filament core itself and not just being part of the short filament population that leads to the enrichment of the galaxy.

When considering galaxies associated with long filaments, $\delta Z_{g}$ varies by $\sim 0.04 \:\text{dex}$ in the full distance range (see Fig. \ref{fig:LongShortdMZRRadial}). We observe a clear increase in enrichment in the core of the long filaments, $d_{skel} \lesssim 1\,\text{cMpc}$, of $\sim 0.03 \:\text{dex}$, however for $d_{skel} \geq 8\,\text{cMpc}$, $\delta Z_{g}$ does not continue falling as it does for the short-filament population. 
Long filaments have been observed to be thinner, with lower-density contrasts to the void, all whilst existing in low-density environments \citep{Castignani2022}. If the density does not change much in the range  $ 4 \lesssim d_{skel} \lesssim 14\,\text{cMpc}$, then the galaxy interaction processes that lead to an increased gas-metallicity will be equally as frequent, leading to similar levels of chemical enrichment independent of $d_{\rm skel}$ over these distances.

For distances in the range $8 \lesssim d_{\rm skel} \lesssim 14.5\,\text{cMpc}$, long-filament galaxies have higher gas metallicities than short-filament galaxies (see Fig. \ref{fig:LongShortdMZRRadial}). While long-filament galaxies have gas metallicities closer to those of the median total MZR and higher enrichment, short-filament galaxies have average gas metallicities comparable to those in the void population. In this distance range, galaxies should be considered to be outside of filaments' influence, as typical filament radii seen in other hydro-dynamical simulations at $z=0$ vary from $\sim 3$ to $5\, \text{Mpc}$ \citep{Galarraga2020}. 

For the value of $\delta Z_{g}$ in long-filament galaxies to not change over this distance range and still be above that of the expected void values, we speculate that galaxies that are clustering around long filaments exist in a more diffused distribution across a larger range of $d_{\rm skel}$, forming new long filaments as they move towards the existing ones. Due to the long filaments' low gravitational potential, the galaxies may not have had the time to accrete towards the core of the long filaments, giving rise to low-density contrasts between the core and the surrounding regions and, as such, higher levels of enrichment in the surroundings relative to the true void. A future analysis into long filaments, the distribution of matter around them and how this changes with redshift will provide an interesting perspective on how these low matter density structures influence galaxies around them over cosmic time, and how the filamentary structures themselves evolve hierarchically.

At a fundamental level, our findings align with the understanding that DM halos experience different dynamical evolution's based on their surrounding densities, which are intricately mapped by the cosmic web. This variance in dynamical timescales affects both the rate at which new stars are formed and the speed with which the intergalactic medium becomes enriched with metals. This bias, related to the density of the environment, has been previously discussed and is known as the Kaiser Bias \citep{Kaiser1984}, or, equivalently, the peak-background split. Considering that the distribution of metals over cosmic time is closely tied to the efficiency of star formation and subsequent feedback processes, it's therefore expected that the MZR is also influenced by the local density, which is distributed anisotropically across the universe. This relationship, influenced by the cosmic web, emphasizes the need to prioritize understanding the interplay between local density and the MZR, as highlighted in recent literature \citep{Donnan2022}. To account for this relationship in our work, we assess the significance of our trends based on $d_{\rm skel}$ and $d_{\rm cluster}$ in Appendix C. 

Moving forward, this work could further illuminate observed scaling laws and their dispersion by examining the spatial correlations between past and current anisotropic clustering. These correlations are indicative of how metals have been dispersed and how their host environments have evolved. By drawing on works like \citet{Musso2018} and \citet{Codis2015}, predictions could be made regarding the clustering of substructures within metal-enriched regions, taking into account how these regions have changed over time and where they are situated within the saddle frame of the cosmic web. A deeper analysis would involve exploring the process of bubble percolation and its contribution to the broader diffusion of metals. This aspect, owing to its non-linear nature, highlights the pivotal role of advanced hydrodynamical simulations, such as HR5, for insights into processes not easily deduced from first principles. 

Lastly, investigating the effects of processes at or near the sub-grid scale on the dispersion of the MZR could shed light on the impact of stochastic elements at these scales. Given that gravitational systems interact across larger and smaller scales, it is plausible that such interactions could significantly influence the distribution and diffusion of metals.

\section{Conclusions}
\label{sec:Concl}

We presented an analysis of the gas metallicity-stellar mass relation (MZR) using the Horizon Run 5 simulation. We studied the effect of the environment (nodes, filaments and voids) on the scatter in the MZR. We further investigated the role of the length of the filaments along with the dynamical state of clusters. We presented how the radial profiles of the deviation from the MZR, $\delta Z_{g}$, change against $d_{\rm skel}$ and $d_{\rm cluster}$ for galaxies near filaments and clusters, respectively, showing more accurately how enrichment changes as one approaches the core of the filaments or the centre of the nodes. From this work, we arrived at the following conclusions.

\begin{enumerate}

    \item Proximity to cluster centers and filament cores leads to a reduction in the average [O/Fe] of $0.08 \, \text{dex}$ in $d_{\rm cluster}$ and $0.03 \, \text{dex}$ in $d_{\rm skel}$ within each environment. 

    We also observe a reduction of $\sim0.4$ with $d_{\rm cluster}$ and $\sim0.1$ with $d_{\rm skel}$ in the average gas fraction with increasing proximity to clusters and filaments. This suggests that the increased rate of Type Ia SNe, which produces iron in these regions, removes further gas from these galaxies. 
    
    \item The cosmic environment plays a crucial role in the vertical scatter of the MZR, producing a substantial vertical deviation from the median of the total distribution of galaxies. The difference in the median metallicities seen between the Node and Void environments is $\sim 0.13 \:\text{dex}$ for a stellar mass range of $9.3 \leq \log(M_{\star}/\text{M}_{\sun})\leq 9.6$. The total scatter due to environment is more prominent at lower stellar masses, mostly in the range $9.3 \leq \log(M_{\star}/\text{M}_{\sun})\leq 10.1$ with only scatter from the Node population existing at $\log(M_{\star}/\text{M}_{\sun}) \geq 10.2$. 
    
    \item Node galaxies show the largest levels of chemical enrichment along with the largest average relative standard deviation of the MZR $\langle \sigma_{Z} \rangle/\langle Z \rangle = 0.596$. Galaxies in void regions, $d_{\rm skel} \geq 8 \text{cMpc}$, demonstrate the lowest levels of chemical enrichment along with the lowest average relative standard deviation of the MZR $\langle \sigma_{Z} \rangle/\langle Z \rangle = 0.264$. We attribute these trends to higher frequencies of galaxy interactions in denser environments.

    \item The scatter in the MZR due to the dynamical state of the node/cluster the galaxies belong to accounts for a $\approx 0.1 \, \text{dex}$ deviation in the gas metallicity between galaxies within relaxed and unrelaxed clusters. Both populations have a positive deviation above the total MZR across the full stellar mass range. The difference in metallicity between the two populations quickly reduces to $\sim 0.05 \, \text{dex}$ at $M_{\star} = 10^{9.6}\,\text{M}_{\sun}$, and they become indistinguishable from each other at high stellar masses, $M_{\star} \geq 10^{10.1}\,\text{M}_{\sun}$.

    \item The Filament population acts as an intermediary population between the Void and the Nodes, showing slight metal enrichment above the median MZR value of the overall distribution of galaxies, with an average relative standard deviation of the MZR $\langle \sigma_{Z} \rangle/\langle Z \rangle = 0.363$, intermediate between the Nodes and the Voids. 

    \item The median residual from the MZR, $\delta Z_{g}$, against $d_{\rm skelshort}$ shows that short filaments have a comparable profile to the overall filament population, reaching median levels of enrichment in the cores, at distances $\lesssim 1$ cMpc, dropping to levels seen in the void with increasing $d_{\rm skelshort}$. Although we still see an increase in $\delta Z_{g}$ in the cores of long filaments, they do not continue to drop to void metal enrichment levels and remain constant at $\delta Z_{g} \approx 0.04$. 

    \item When considering galaxies with fixed $M_{\star}$, the gas metallicity, $Z_{g}$, has a weak correlation with the SFR within the filaments and voids at redshift $z=0.625$ in the simulation. A stronger correlation between $Z_{g}$ and SFR is predicted for the node galaxy population, as galaxies within more virialised nodes have, on average, higher $Z_{g}$ and lower SFRs; this causes a larger tilt angle of the predicted FMR plane of the node galaxy population.
    
\end{enumerate}

    Although the vertical scatter in the MZR is caused by a plethora of galactic and cosmic processes, we have begun to unravel the effects associated with just one of these contributors. In the context of the cosmic web, gas metallicity, along with other galaxy properties, are of great importance to delve into in the coming years to further build on our understanding of the incomplete story of galaxy evolution. Upcoming surveys like the Dark Energy Spectroscopic Instrument \citep{DESI2016,DESI2023}, the Prime Focus Spectrograph (PFS) \citep{Greene2022} and Euclid \citep{Racca2016, Euclid2022}, will allow us to probe the observed large scale structure in nature more easily, and as such explore the effects presented in this paper in greater detail. This work explores the issue from a simulation perspective to compare and contrast our understanding of the universe, building upon great work to lay a foundation for further studies into galaxy evolution in a cosmological context.
    In the future, we aim to explore how metallicity can be a useful tool in understanding the redshift evolution of filaments and the galaxy populations surrounding them. Building upon Fig. \ref{fig:LongShortdMZRRadial}, we hope to begin to explain this flattening of $\delta Z$ around longer filaments and how this trend changes, not only in redshift but also when considering a larger number of filament length bins. 
    It is also of future interest to probe the true multi-scale nature of the cosmic web. By looking to identify more detailed cosmic structures over a larger range of sizes, one could begin to disentangle the levels of hierarchical significance of each spatial range, shedding light on whether the large or small-scale filamentary systems dominate over the observed trends.

\section*{Acknowledgements}

We thank the reviewer for their insight into the paper and their suggestions which elevated the work we have presented. We acknowledge the support of STFC (through the University of Hull's Consolidated Grant ST/R000840/1) and ongoing access to {\tt viper}, the University of Hull High-Performance Computing Facility. BKG and FV thank the National Science Foundation (NSF, USA) under grant No. PHY-1430152 (JINA Center for the Evolution of the Elements).  BKG acknowledges the support of the European Union’s Horizon 2020 research and innovation programme (ChETEC-INFRA – Project no. 101008324). We acknowledge the support of computing resources at the Center for Advanced Computation (CAC) at KIAS. AS is supported by a KIAS Individual Grant PG080901 at the Korea Institute for Advanced Study (KIAS). SEH was partly supported by the Korea Astronomy and Space Science Institute grant funded by the Korea government (MSIT) (No. 2024186901) and the grant funded by the Ministry of Science (No. 1711193044). J.L. is supported by the National Research Foundation of Korea (NRF-2021R1C1C2011626). This work was also partially supported by the National Research Foundation of Korea (NRF) grant funded by the Korea government (MSIT, 2022M3K3A1093827). Y.K. is supported by Korea Institute of Science and Technology Information (KISTI) under the institutional R\&D project (K24L2M1C4). This work is partially supported by the grant \href{https://www.secular-evolution.org}{Segal} ANR-19-CE31-0017 of the French Agence Nationale de la Recherche. 

\section*{Data Availability Statement}

The data underlying this work is able to be shared upon reasonable request.

\bibliographystyle{mnras}
\bibliography{references}

\appendix

\section{Comparing Methodologies of Skeleton Computation with T-ReX}

\begin{figure*}
    \centering
    \includegraphics[scale=0.6]{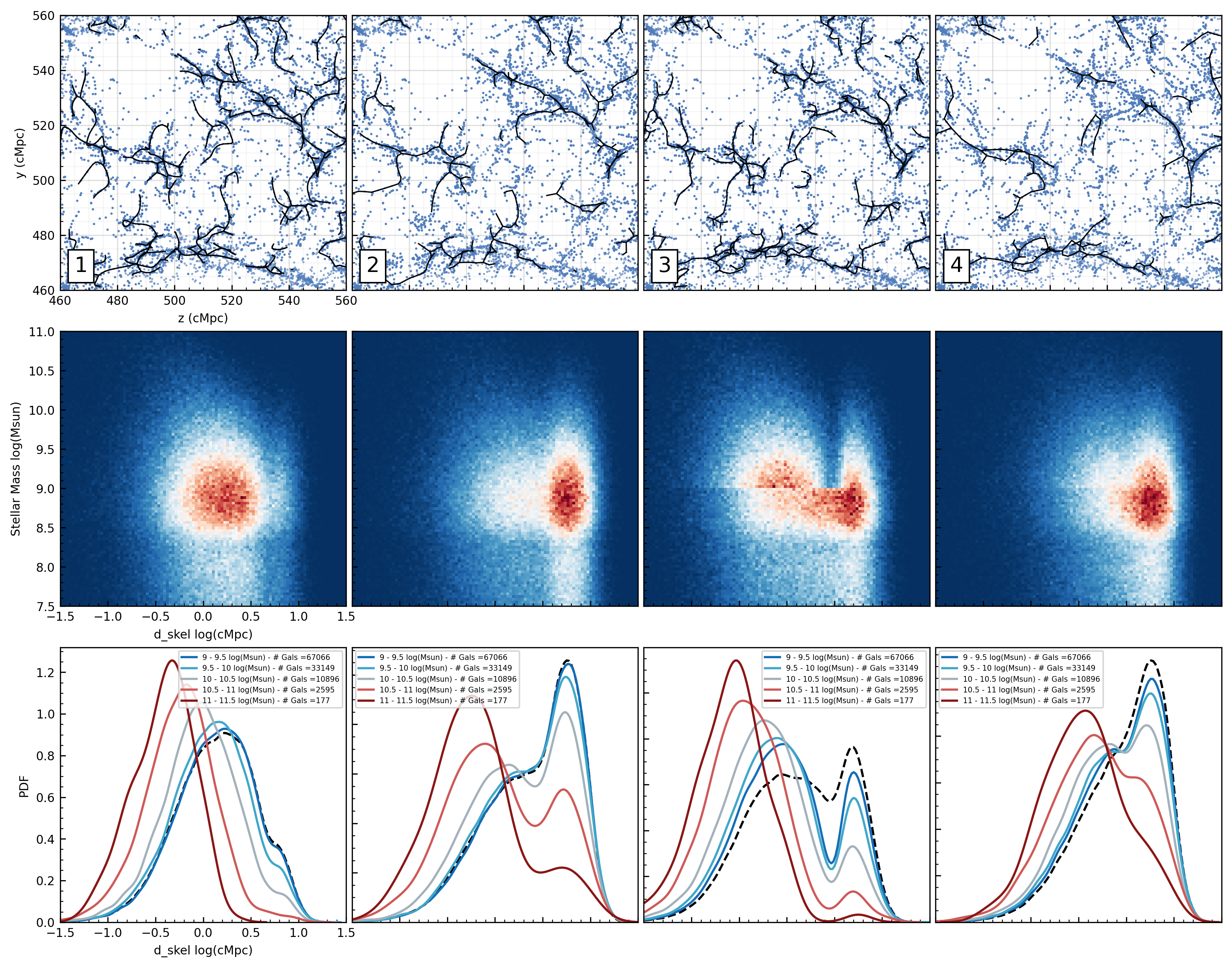}
    \caption{A comparison of four separate methodologies, each providing {\tt T-ReX} with a different galaxy distribution. The top row of plots shows 100x100x10 cMpc slices in HR5, each point representing a galaxy sub-halo. These are overlayed with the skeleton black lines, computed using {\tt T-ReX} for the corresponding galaxy distribution. The second row shows 2d histograms of Stellar Mass against the calculated radial distances, $d_{\rm skel}$, for each methodology. No color bar is included, as the broad shapes of these histograms are the main result of this analysis. The final row shows the 1d PDF of the distance values for each methodology, split into five stellar mass bins. The black dotted line shows the 1d PDF for the total distribution of galaxies irrelevant to stellar mass. Note that none of these plots show the final skeleton that is used in the analysis; they represent four methodologies from which a starting point can be found.}
    \label{fig:figa1}
\end{figure*}

To arrive at our methodology of skeleton computation using {\tt T-ReX}, we tested four separate methodologies that give {\tt T-ReX} alternate galaxy distributions based on cuts in stellar mass and taking a random selection of galaxies. We could then produce radial distance values for each galaxy using these skeletons. Observing their PDFs, we could select the most robust methodology, correctly re-producing expected profiles and distributions in $\ell_{\rm fil}$ and $d_{\rm skel}$. Fig. \ref{fig:figa1} shows the overall comparison of the four methodologies. To begin the analysis, we created four galaxy catalogues to give to {\tt T-ReX}. Table \ref{tablea1}. outlines these galaxy distributions.

\begin{center}
\begin{table}
\begin{tabular}{ |c|c|c|c|c|c| }
 Method & $M_{\star}$ Cut & Rand \% of Total & $\Lambda$ & l & $\sigma$\\ 
 \hline
 1 & None & 100\% & 5.0 & 10.0 & 3.0\\  
 \hline
 2 & None & 20\% & 1.0 & 5.0 & 1.0 \\
 \hline
 3 & $\geq 10^{9}$ & 100\% & 1.0 & 5.0 & 1.0 \\
 \hline
 4 & $\geq 10^{9}$ & 20\%  & 1.0 & 1.0 & 1.0 \\
\hline
\end{tabular}
\caption{The parameter choices and overview of each of the 4 methodologies.}
\label{tablea1}
\end{table}
\end{center}

Method 1 gives {\tt T-ReX} the total galaxy distribution, only removing galaxies flagged as 'impure' in the galaxy catalogue; this represents the idea that the filament finder will benefit from receiving as much information as possible such that it will produce the most detailed and accurate skeleton. Method 2 is an extension of this idea in which we take a random $20\%$ selection of the distribution. This tests whether a random selection of the total distribution can correctly map the structure. Method 3 does not use this random selection but instead takes a stellar mass cut. As mentioned in the methodology section of the paper, this method is common in other studies as it has been shown that high stellar mass galaxies are more commonly found within the cosmic structure and, therefore, are good tracers. For this method, we take the total distribution of galaxies after removing the `impure' galaxies and remove all galaxies with  $M_{\star} \leq 10^{9}$. Method 4 repeats this cut and then again takes a random $20\%$ sample of the distribution.

Each distribution is then handed to {\tt T-ReX} to compute a corresponding skeleton for each. {\tt T-ReX} was specifically tuned for each galaxy distribution depending on the number of galaxies selected; specifics of this tuning can be seen in Table \ref{tablea1}. Using the skeletons, we then calculate each galaxy's perpendicular distance to its closest edge on each of them, giving us $d_{\rm skel1}$, $d_{\rm skel2}$, $d_{\rm skel3}$, $d_{\rm skel4}$. From each set of distances, we can then produce 2d histograms against stellar mass and 1d PDFs of these values, with five equally-spaced, increasing stellar mass bins from $10^{9}-10^{11.5} M_{\odot}$.

Starting in the top row of Fig. \ref{fig:figa1}, we see the computed skeleton associated with each methodology overlayed onto the total galaxy distribution in the snapshot. Immediately, we can observe that Methods 1 and 3 produce skeletons with the highest filament density, with many filaments being 'found' in the most dense clusters of the distribution. On the other hand, Method 2 and 4 display skeletons with lower filament density, with Method 4's skeleton being the least dense. Due to the nature of visual inspection and how it can lead to differences in interpretation from peer to peer, it is not easy to base a decision on this purely. As such, we turn our attention to the middle row of plots.

On the middle row, we see the 2d histograms of $d_{\rm skel}$ against stellar mass. These plots are where very clear differences begin to arise. The main result from these plots comes from comparing methods 1 and 2 with 3 and 4. Methods 3 and 4 present a clear, non-physical line in the histogram at $10^{9} M_{\odot}$, the exact stellar mass value that we took the cut. Galaxies above this line have been used in the skeleton computation; galaxies below this line have not and have been placed onto the existing skeleton for their value of $d_{\rm skel}$ to be calculated. This demonstrates that $d_{\rm skel}$ values depend on whether the galaxy in question has been used in the skeleton computation. Therefore, to have a coherent and usable set of distances, we must be consistent in our calculation for all galaxies being used in the analysis, leading to method 1 or 2 being our best option. From the 2d histograms of methods 1 and 2, we see a clear difference in the distance at which the main peak in the distribution exists. For method 1, the peak lies in the $0-0.5$ log(cMpc) range, whilst in method 2, the peak lies in the $0.5 - 1$ log(cMpc) range. We can now turn to the 1d PDFs to further inform our findings.

All methods, apart from method 1, show a clear bimodality in $d_{\rm skel}$ with the first peak sitting in the range of $-0.5 - 0 \text{ log(cMpc)}$, and the second peak sitting at $0.8 \text{ log(cMpc)}$. The main difference between methods 1 and 2 in this space is that the bimodality exists in method 2 and does not exist in method 1. The only difference in these methodologies is that method 2 uses $20\%$ of the galaxy distribution in the skeleton computation, which produces a much less overpopulated skeleton whilst still tracing the large-scale structures. We can explain the uni-modality in method 1's PDF by considering the overpopulation in the skeleton. Due to the large number of galaxies given to {\tt T-ReX} and the parameters chosen, the algorithm creates a very large number of filaments such that most galaxies will be found to lie on one. This means that we end up seeing a prominent single peak, as the skeleton is so dense all the galaxies are found to lie --nearby-- to filaments. The second peak at $0.8 log(cMpc)$, seen in the PDF for method 2, can, in turn, represent the average distance between the identified skeleton and galaxies outside of it. With these final two methods, we also looked at the distribution of filament lengths, and as expected, method 1 produced far too many short filaments, leading to a bimodal distribution, which is not expected in literature, see \citealt{Galarraga2023_2}. Method 2 produced a PDF that was closer to what was expected. From this, we settled that method 2 is the best of the four methods we present here, acting as a good starting point for the final skeleton.

To create our final skeleton, we slightly tuned the parameters around the provided values seen for Method 2 until the skeleton we produced had $d_{\rm skel}$ and $\ell_{\text{fil}}$ PDFs that agreed with what has been used in existing work, \citep{Donnan2022, Bulichi2023} and \citep{Galarraga2023_2} respectively. These final parameters are; $\Lambda = 5$, $l = 2$ and $\sigma = 2$.

\begin{figure}
    \centering
    \includegraphics[scale=0.65]{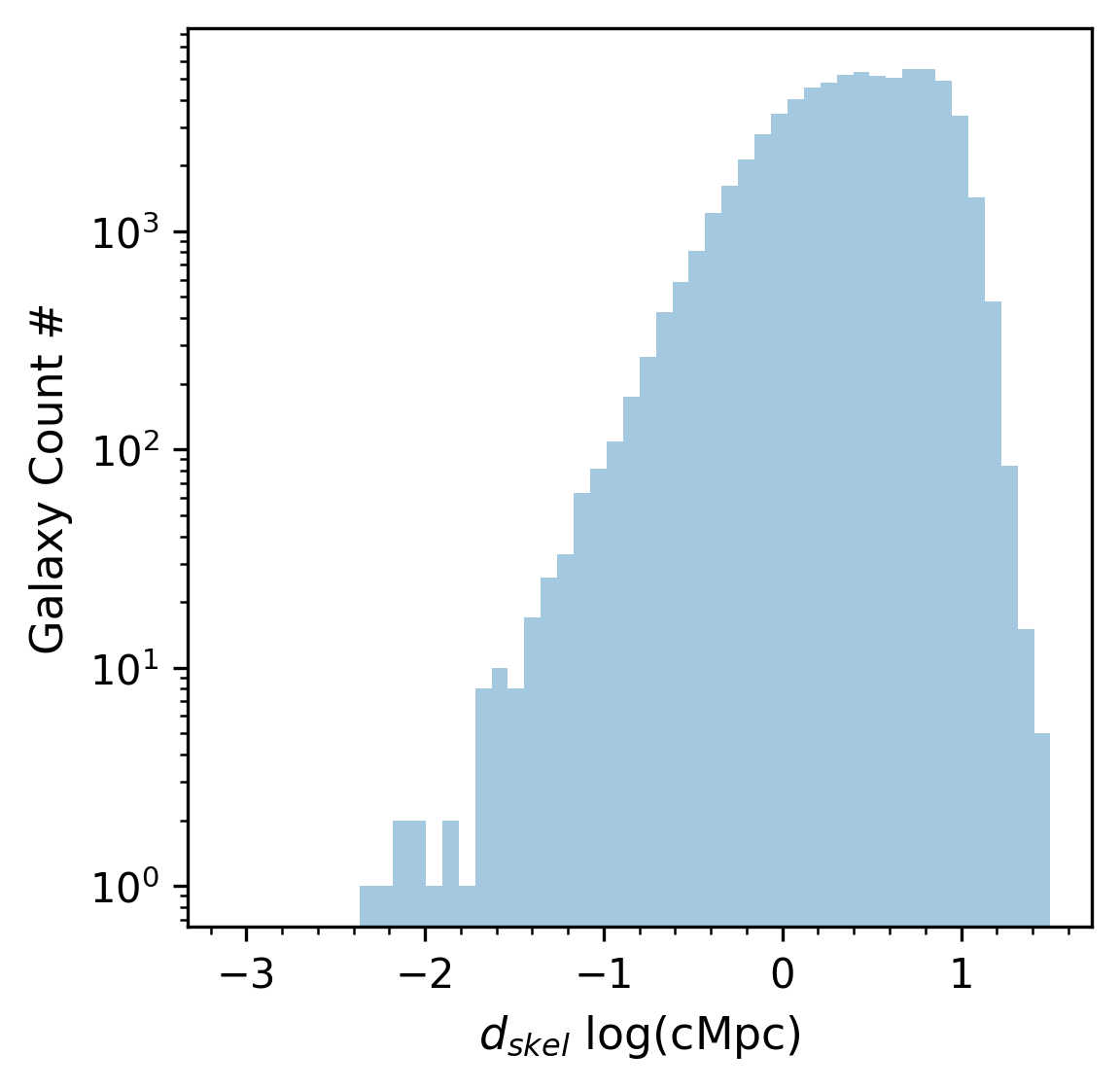}
    \caption{The total distribution of $d_{\rm skel}$ across 50 equally sized logarithmic bins.}
    \label{fig:figa2}
\end{figure}

\begin{figure}
    \centering
    \includegraphics[scale=0.65]{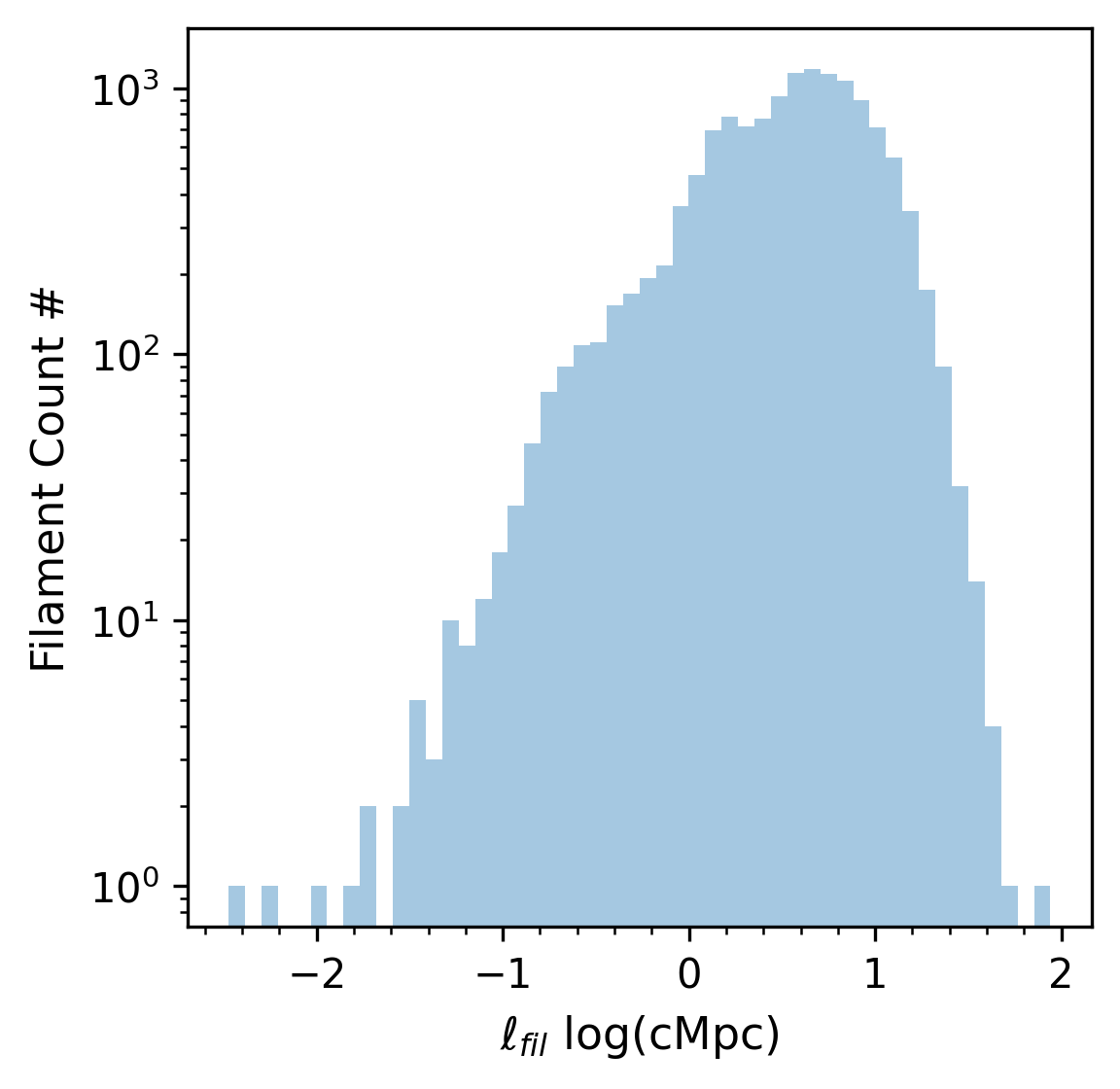}
    \caption{The total distribution of $\ell_{\text{fil}}$ across 50 equally sized logarithmic bins.}
    \label{fig:figa3}
\end{figure}

Fig. \ref{fig:figa2} and Fig. \ref{fig:figa3} show the distributions for $d_{\rm skel}$ and $\ell_{\text{fil}}$ from the final skeleton, giving us confidence that our skeleton is correctly mapping large scale structure in accordance to other previous studies carried out in the field using similar methodologies and filament finders.

\section{Dynamical State of Clusters in HR5}

\begin{figure*}
    \centering
    \includegraphics[scale=0.65]{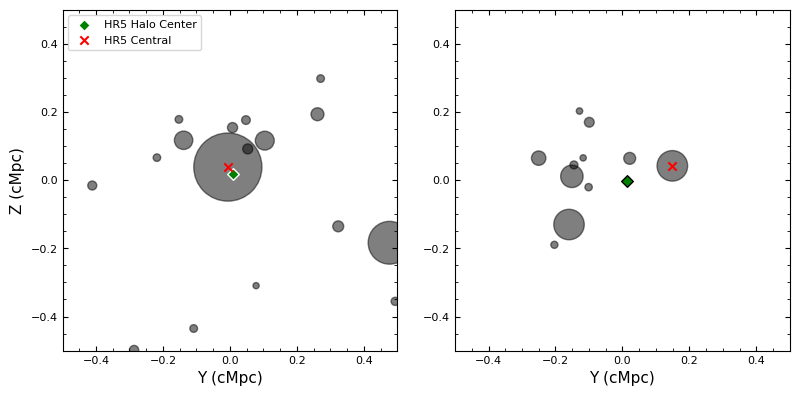}
    \caption{Two separate clusters from HR5's Halo catalogue in halo-centric coordinates. The shaded grey circles are galaxies in the cluster, scaled to the galaxy's stellar mass to demonstrate the distribution of galaxy stellar mass in the plot. The red cross represents the central galaxy of the halo, and the green diamond represents the FoF center of mass in the halo, both defined by the FoF algorithms and {\tt PGalF} in HR5. The left plot shows the situation with minimal separation between the FoF center of mass and the identified Central galaxy of the halo, representing a relaxed cluster by our definition. The right plot shows the opposite situation with a $\sim 130 \text{ ckpc}$ separation between the two positions. This represents our unrelaxed cluster situation.}
    \label{fig:UnrelaxedvsRelaxed}
\end{figure*}

To more clearly demonstrate the offset that exists between relaxed and unrelaxed clusters in HR5, Fig. \ref{fig:UnrelaxedvsRelaxed} shows a galaxy cluster that lies in each of our populations based on $\Delta_{r}$. It is clearly shown that in the right plot, the offset between the central galaxy and FoF center of mass is large enough such that our $d_{\rm cluster}$ values are measuring to the incorrect position for what we wish to probe. In the left plot, we can see a virialized and settled-down cluster with a much smaller offset, meaning the $d_{\rm cluster}$ values associated with these clusters are meaningful in our analysis.

\section{Correcting against the expected over-densities of the cosmic web}

As is known from existing studies, the cosmic web and measurements relative to it are strongly correlated with over densities. As such, when providing an analysis referring to galaxy properties as a function of, or relative to, $d_{\rm skel}$, it is reasonable to assess how significant this parameter is in relation to these known trends with changing densities. A simple approach is to calculate a value of local density for each galaxy and using these values repeat the analysis using a limited density range. One can imagine here that if the trends vanished after limiting density, the over-densities associated with the large-scale structure are the main cause of the observed trends. Conversely, if the trends remain, significance can be placed in the other parameter in question, $d_{\rm skel}$. This is carried out in \citealt{Donnan2022} in which it is shown that their trends with $d_{\rm skel}$ persist once a density cut is taken, and we choose to implement a similar method. 

\begin{figure}
    \centering
    \includegraphics[scale=0.65]{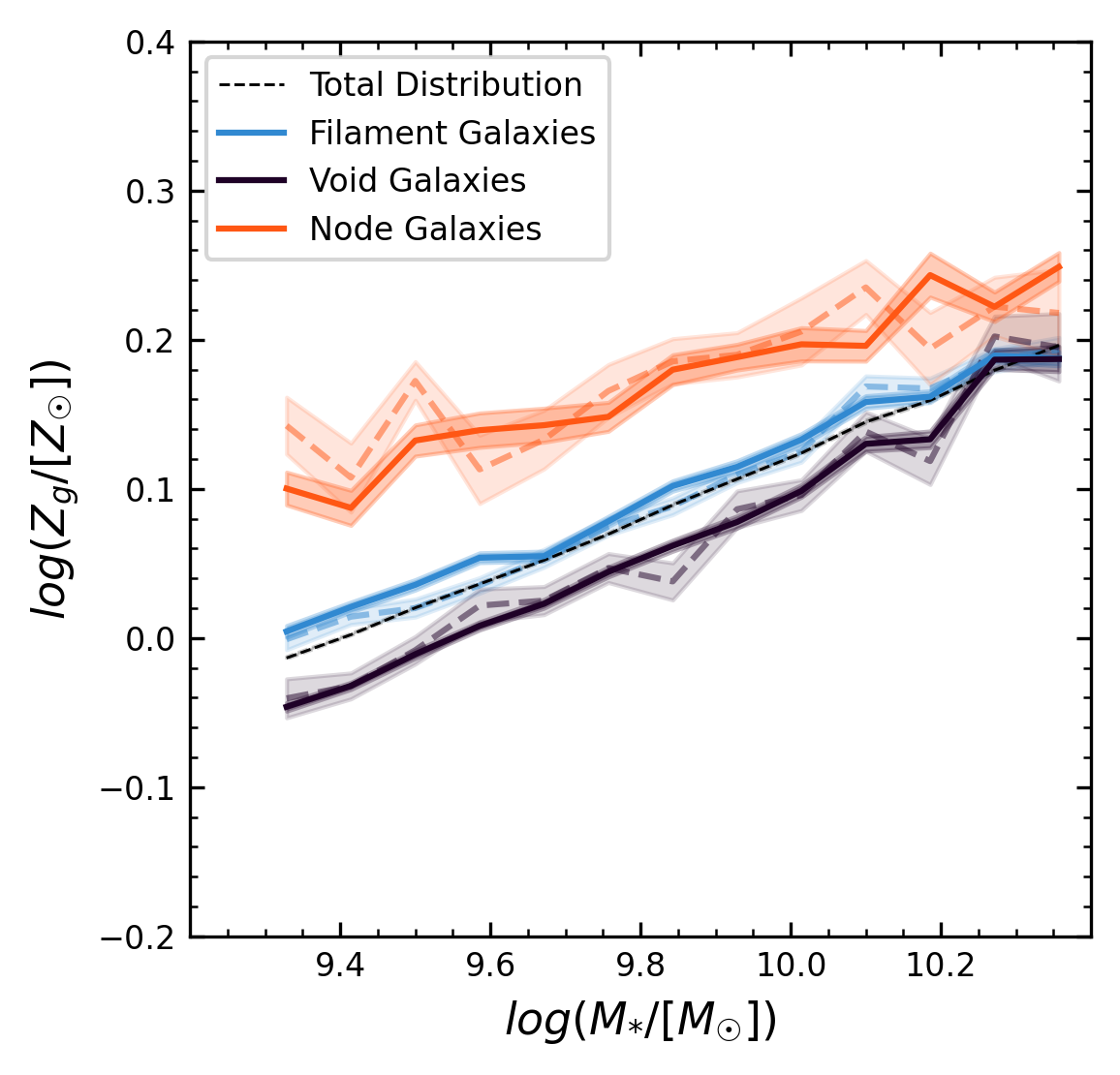}
    \caption{The galaxy stellar mass-gas phase metallicity relation in different environments as seen in Fig. \ref{fig:MZRScatterNFV} The node population is shown in orange, the filament population in blue, and the void population in dark purple. The black dashed line shows the median MZR for the total galaxy distribution. The coloured shaded regions show the standard error on the median for each of the three environments. The lighter dashed lines represent the MZR associated with the environments within a small range of $\rho_{\rm G}$ of $0.052 \pm 0.021\, \text{cMpc}^{-3}$.}
    \label{fig:DensityMZR}
\end{figure}

We calculate local densities of each galaxy, $\rho_{G}$, as simple spherical hats, taking the number density of galaxies within a 3 cMpc sphere around the selected galaxy.
This provided a median $\rho_{\rm G}$ of $0.053 \, \text{cMpc}^{-3}$. Following \citealt{Donnan2022}, we then take our cut around this median. In their work, they choose to take $\pm 0.32 \sigma$ around their median. However, due to our large statistical sample of galaxies, we choose a stricter cut of $\pm 0.15 \sigma$, whilst maintaining a statistically significant number of galaxies. From Fig. \ref{fig:DensityMZR}, we can see that with a constrained density cut, the trends with environment still persist and as such $d_{\rm skel}$ itself is significantly contributing to these trends, above the expected trend with over-densities in the cosmic web. This result is in agreement with what is shown for the same analysis in \citep{Donnan2022}.

\end{document}